\documentclass[twocolumn,superscriptaddress,
amsmath,amssymb,aps,pra]{revtex4-1}

\usepackage{graphicx}
\usepackage{physics}
\usepackage{comment}
\usepackage{color}
\usepackage{dcolumn}
\usepackage{bm}

\usepackage[colorlinks,urlcolor=blue,citecolor=blue,linkcolor=blue]{hyperref}

\renewcommand{\k}{{\bf k}}

\newcommand{\q}{{\bf q}}
\newcommand{\Q}{{\bf Q}}

\newcommand{\0}{{\bf 0}}

\newcommand{\ef}{E_F}

\newcommand{\nn}{\nonumber}
\newcommand{\beq}{\begin{equation}}
\newcommand{\eeq}{\end{equation}}

\newcommand{\area}{\mathcal{A}}

\DeclareMathOperator{\im}{\mbox{Im}}
\DeclareMathOperator*{\Simiq}{\simeq}
\newcommand{\Frac}[2]{\displaystyle\frac{#1}{#2}}

\begin{document}

\title{Crossover from exciton polarons to trions in doped two-dimensional semiconductors at finite temperature}

\author{Antonio Tiene}
\thanks{A. T. and B. C. M. contributed equally to this work.}
\affiliation{Departamento de F\'isica Te\'orica de la Materia
  Condensada \& Condensed Matter Physics Center (IFIMAC), Universidad
  Aut\'onoma de Madrid, Madrid 28049, Spain}
\affiliation{School of Physics and Astronomy, Monash University, Victoria 3800, Australia}

\author{Brendan C. Mulkerin}
\thanks{A. T. and B. C. M. contributed equally to this work.}
\affiliation{School of Physics and Astronomy, Monash University, Victoria 3800, Australia}
\affiliation{ARC Centre of Excellence in Future Low-Energy Electronics Technologies, Monash University, Victoria 3800, Australia}

\author{Jesper~Levinsen}
\affiliation{School of Physics and Astronomy, Monash University, Victoria 3800, Australia}
\affiliation{ARC Centre of Excellence in Future Low-Energy Electronics Technologies, Monash University, Victoria 3800, Australia}

\author{Meera M. Parish}
\affiliation{School of Physics and Astronomy, Monash University, Victoria 3800, Australia}
\affiliation{ARC Centre of Excellence in Future Low-Energy Electronics Technologies, Monash University, Victoria 3800, Australia}

\author{Francesca Maria Marchetti}
\affiliation{Departamento de F\'isica Te\'orica de la Materia
  Condensada \& Condensed Matter Physics Center (IFIMAC), Universidad
  Aut\'onoma de Madrid, Madrid 28049, Spain}

\date{\today}

\begin{abstract}
    We study systematically the role of temperature in the optical response of doped two-dimensional semiconductors. By making use of a finite-temperature Fermi-polaron theory, we reveal a crossover from a quantum-degenerate regime with well-defined polaron quasiparticles  to an incoherent regime at high temperature or low doping where the lowest energy ``attractive'' polaron quasiparticle is destroyed, becoming subsumed into a broad trion-hole continuum. 
    We  demonstrate that the crossover is accompanied by significant qualitative changes in both absorption and photoluminescence. In particular, with increasing temperature (or decreasing doping), the emission profile of the attractive branch evolves from a symmetric Lorentzian to an asymmetric peak with an 
    exponential tail involving trions and recoil electrons at finite momentum. We discuss the effect of temperature on the coupling to light for structures embedded into a microcavity, and we show that there can exist well-defined polariton quasiparticles even when the exciton-polaron quasiparticle has been destroyed, where the transition from weak to strong light-matter coupling can be explained in terms of the polaron linewidths and spectral weights.
\end{abstract}

\maketitle

\section{Introduction}
\label{sec:intro}
The notion of a quantum impurity, where a particle is dressed by excitations of a quantum gas, was first introduced by Landau in 1933~\cite{Landau1933} to describe the behavior of conduction electrons in a dielectric medium. The properties of this dressed impurity (or polaron), such as mobility or effective mass, are changed with respect to those of an isolated impurity, leading to strong modifications of, e.g., electrical and thermal transport. Despite nearly a century of work, the polaron problem continues to attract significant interest, with noticeable realizations in ultracold atomic gases~\cite{Massignan_RPP2014,Levinsen2Dreview,Scazza2022}, $^3$He impurities in $^4$He~\cite{Bardeen1967}, and the Kondo effect generated by localized magnetic impurities in a metal~\cite{Kondo1964}, just to cite a few examples. 

Recently, the absorption and emission spectra of doped two-dimensional (2D) transition metal dicalchogenide (TMD) monolayers~\cite{Sidler_NatPhys_2017,Goldstein_JCP2020,Koksal_PRR2021,Liu_NatComm2021,Xiao_JPCL2021,Zipfel_PRB2022} have been interpreted in terms of a Fermi-polaron model~\cite{Suris2003correlation,Efimkin2017,Rana_PRB2020,Rana_PRB2021,Rana_PRL2021,Efimkin_PRB_2021}. Here, the optically generated exciton is dressed by excitations of a 2D Fermi gas of charge carriers (electrons or holes) induced by either gating or natural doping of the TMD monolayer. This leads to the formation of ``attractive'' and ``repulsive'' Fermi polarons, 
where the exciton attracts or repels the surrounding charge carriers, 
respectively. 
Furthermore, when such a TMD monolayer is embedded in a microcavity~\cite{Sidler_NatPhys_2017,Chakraborty_NanoLett2018,Emmanuele_NatComm2020,Koksal_PRR2021,Lyons_NatPh2022},
the strong coupling between light and matter can lead to the formation of Fermi polaron polaritons. 

The intense recent activity on this topic is motivated by several features of  exciton-polaron quasiparticles. 
The dressing of %
polaritons by a Fermi gas can boost their pairwise interactions or non-linearities~\cite{Tan2020,Emmanuele_NatComm2020}, which can potentially be exploited to reach the
polariton blockade regime~\cite{Kyriienko_PRL2020}. Another interesting aspect of this configuration is the possibility of manipulating dressed excitons or polaritons by external electric and magnetic fields~\cite{Efimkin2018,Cotlet_PRX2019,Smolenski_PRL2019,Chervy_PRX2020}. Finally, exciton polarons can be used as a probe of strongly correlated states of electrons, such as Wigner crystals in a TMD monolayer~\cite{Smolenski_Nature2021,Shimazaki_PRX2021}, quantum Hall physics in TMD monolayers~\cite{Smolenski_PRL2019}, fractional quantum Hall states in proximal graphene layers~\cite{Propert_NanoLett2022}, and correlated Mott states of electrons in a Moir\'e superlattice~\cite{Schwartz_science2021}.

The theoretical analysis of the Fermi polaron in 2D semiconductors has so far mainly focused on the zero-temperature limit~\footnote{
While the results in Refs.~\cite{Baeten_PRB2015,Rana_PRB2020} have been derived at finite temperature, the former only considered infinite exciton mass, while the latter did not analyze the consequences of temperature on the polaron description.}. 
However, a natural question to ask is how the system changes with temperature, since this has been shown to strongly modify the nature of the Fermi polaron in cold-atom experiments~\cite{Yan2019a}.
In particular, there is a competing picture to that of the Fermi polaron  based on 
few-body complexes such as excitons and trions, which provides a description for the photoluminescence of the attractive branch in doped semiconductors at finite temperature~\cite{Bronold2000,Carbone_JCP2020,Zhumagulov_JCP2020}.  

In this paper, we use a finite-temperature variational approach 
developed in the context of cold atoms~\cite{Liu2019} to reveal the important role that temperature plays in the exciton-polaron problem. 
In particular, we demonstrate that, when the temperature becomes large compared to the Fermi energy of charge carriers,
the attractive Fermi polaron merges with a continuum of states comprised of a bound trion and unbound Fermi-sea-hole---the so-called trion-hole continuum.
Here, the attractive branch ceases to be a well-defined quasiparticle and it crosses over into an incoherent regime dominated by the trion-hole continuum. We discuss the implications of this crossover for the existence of polariton quasiparticles when the semiconductor is strongly coupled to light in a microcavity.
By contrast, we find that temperature does not qualitatively change the nature of the repulsive branch at typical doping levels, only its polaron properties.

We find
that the destruction of the attractive polaron quasiparticle 
has little effect on the energy and spectral weight, but it strongly modifies the linewidth and 
the overall shape of the attractive peak in both absorption and photoluminescence. 
In particular, we observe that 
the attractive branch %
evolves from a symmetric Lorentzian shape to a strongly asymmetric profile with an exponential tail below the trion energy. %
In the latter 
trion-dominated regime, our theory becomes perturbatively exact since it corresponds to the lowest order term in a 
quantum virial expansion~\cite{finiteTshort}.  
Note that, even though our model is formulated to describe doped 2D semiconductors, our results can be easily generalized to 2D atomic Fermi gases~\cite{Koschorreck2012,Zhang2012,Oppong2019}, and they can straightforwardly be extended to the three-dimensional case.

The paper is organized as follows: In Sec.~\ref{sec:model} we introduce the model describing excitons and electrons in a doped TMD monolayer. In Sec.~\ref{sec:polaron} we describe the variational approach for impurities at finite temperature, we relate it to the $T$-matrix approach, and we define absorption and photoluminescence. In Sec.~\ref{sec:results} we illustrate the results for optical absorption and photoluminescence, comparing the case of finite temperature against the zero temperature limit. We also carry out a perturbatively exact quantum virial expansion to relate the results of this work with those of the accompanying paper~\cite{finiteTshort}. In Sec.~\ref{sec:strong-c} we extend the formalism and results to the case where the doped TMD monolayer is embedded into a microcavity to allow for strong light-matter coupling. Finally, in Sec.~\ref{sec:concl} we conclude and present future perspectives of this work.

\section{Model}
\label{sec:model}
We consider the following model Hamiltonian describing a doped 2D semiconductor such as a TMD monolayer:
\begin{subequations}
\label{eq:Hamiltonian}
\begin{align}
    \hat{H} &= \hat{H}_0 + \hat{H}_{0X} +\hat{H}_{int}\; ,
    \\
    \hat{H}_0&=
    \sum_{\k} (\epsilon_{\k} -\mu)\hat{c}^{\dag}_{\k}\hat{c}^{}_{\k}\; ,
 \\
    \hat{H}_{0X}&=
    \sum_{\k} 
    \epsilon_{X\k}\hat{x}^{\dag}_{\k}\hat{x}^{}_{\k}\; ,
    \\
    \label{eq:H_Xe}
    \hat{H}_{int}&=
   -\frac{v}{\area}\sum_{\k\k'\q}\hat{x}^{\dag}_{\k+\q}\hat{c}^{\dag}_{\k'-\q} \hat{c}^{}_{\k'}\hat{x}^{}_{\k} \; ,
\end{align}
\end{subequations}
where $\area$ is the system area (throughout this paper we set $\hbar = k_B= 1$). The Hamiltonian $\hat{H}_0$ describes the fermionic medium-only part, i.e., a Fermi gas of either electrons or holes, in terms of the fermionic operator $\hat{c}^{\dag}_{\k}$ that creates a charge carrier with mass $m$ and dispersion $\epsilon_{\k}= |\k|^2/2m \equiv k^2/2m$. For concreteness, we consider the specific case where the charge carriers are electrons, but our results can be trivially extended to the case of hole doping. For simplicity,
we treat electrons 
as non-interacting, an assumption which becomes exact in the high-temperature, low-doping limit~\cite{finiteTshort}. %
We furthermore ignore the spin/valley degree of freedom such that all the electrons are spin polarized but distinguishable from the electron within the exciton (see Ref.~\cite{Tiene_PRB2022} for a theoretical analysis of the case where the charge carriers are indistinguishable). This minimal model is reasonable for describing the polaron and trion physics in TMD monolayers, such as MoSe$_2$ monolayers~\cite{Sidler_NatPhys_2017,Efimkin2017,Efimkin_PRB_2021}.

The medium is described within the grand canonical ensemble, where the chemical potential $\mu$ is
related to the excess electron %
density $n=N/\area$ by 
\begin{align}
    \mu &= T \ln \left(e^{\beta E_F} -1 \right)\;, & E_F &=\frac{2\pi}{ m} n\; ,
\end{align}
where $E_F$ is the Fermi energy and $\beta = T^{-1}$ the inverse temperature.
While we use a grand canonical formalism for the medium, it is convenient to consider only a single excitonic impurity in the canonical formalism~\cite{Liu2019}, as in $\hat H_{0X}$. Formally, this corresponds to considering uncorrelated excitons, and this description is thus appropriate for a low density of excitons, i.e., the quantum impurity limit. Following previous works~\cite{Sidler_NatPhys_2017,Efimkin2017,GlazovJCP2020},  we treat the exciton as a structureless boson, with corresponding creation operator $\hat{x}^{\dag}_{\k}$, mass $m_X$ and free-particle dispersion $\epsilon_{X\k}=k^2/2m_X$. This is because in TMD monolayers the optically generated exciton is tightly bound and its binding energy is larger than all other relevant energy scales of the problem. Note that, throughout this paper, energies are measured with respect to that of the exciton at rest.

The $H_{int}$ term~\eqref{eq:H_Xe} describes the short-range attractive 
interaction between electrons %
and excitons, which can be approximated as a contact interaction~\cite{Fey_PRB_2020,Efimkin_PRB_2021} with coupling strength $v>0$ up to a high-momentum cutoff $\Lambda$. 
The renormalization of such an interaction can be carried out by
relating the bare parameters to the trion binding energy $\varepsilon_T$:
\begin{equation}
    \frac{1}{v} = \frac{1}{\area}\sum_{\k}^{\Lambda} \frac{1}{\varepsilon_T+\epsilon_{X\k} + \epsilon_{\k}} \; .
\label{eq:trion_binding-en}    
\end{equation}
Note that all our results are independent of cutoff since we formally take $\Lambda \to \infty$. 

Our formalism is applicable to a general quantum impurity in a 2D Fermi gas, including both ultracold atomic gases and doped semiconductors. 
To be concrete, in the following we will consider the experimentally relevant case of electron-doped MoSe$_2$ monolayers, where $\varepsilon_T\simeq 25$ meV~\cite{Sidler_NatPhys_2017,Zipfel_PRB2022} and $m_X/m=2.05$~\cite{Kormanyos_2DMat_2015}. In this system one can readily achieve doping densities $E_F$ in the range of 0--40~meV~\cite{Sidler_NatPhys_2017}.

\section{Finite-temperature ansatz}
\label{sec:polaron}
At finite temperature, one can employ the variational approach for impurity dynamics developed in Ref.~\cite{Liu2019} in the context of the Fermi polaron problem in ultracold atomic gases. This variational approach has been successfully used to model dynamical probes such as Ramsey spectroscopy~\cite{Cetina2016,Liu2019} and Rabi oscillations~\cite{Scazza2017,Adlong2020}, as well as static thermodynamic properties such as the impurity contact~\cite{Yan2019a,Liu2020prl}. For completeness, we review the approach here, formulating it for the 2D semiconductor problem. We consider the case of zero impurity momentum, relevant for evaluating absorption and emission. The generalization to finite impurity momentum is discussed in Appendix~\ref{app:finiteQ}. 

The starting point of the variational approach~\cite{Liu2019} is the time-dependent impurity operator that approximates the exact  operator in the Heisenberg picture, $\hat{x}_{\0}^{}(t) = e^{i\hat{H}t} \hat{x}_{\0}^{} e^{-i\hat{H}t}$. We choose the form
\begin{equation}
    \hat{x}_{\0}^{}(t) \simeq \varphi_{0}^{}(t) \hat{x}_{\0}^{} + \frac{1}{\area} \sum_{\k,\q} \varphi_{\k\q}^{}(t)  \hat{c}_{\q}^{\dag} \hat{c}_{\k}^{} \hat{x}_{\q-\k}^{}\; ,
\label{eq:var_imp-op}    
\end{equation}
which is written in terms of the time-dependent variational coefficients $\varphi_{0}^{}(t)$ and $\varphi_{\k\q}^{}(t)$. The truncated form of this operator is similar to that of the Chevy ansatz~\cite{Chevy2006} employed for the zero-temperature state, and describes an impurity dressed by a single excitation of the fermionic medium.

The time-dependent exciton operator~\eqref{eq:var_imp-op} does not coincide with the exact solution of the Heisenberg equation of motion, and thus we determine the variational coefficients by minimizing the error function
\begin{equation}
    \Delta (t) = \langle \hat{e}(t) \hat{e}^{\dag}(t) \rangle_{\beta} \equiv \Tr [\hat{\rho}_0 \hat{e}(t) \hat{e}^{\dag}(t)]\; ,
\label{eq:error}    
\end{equation}
with respect to $\varphi_{0}^{*}(t)$ and $\varphi_{\k\q}^{*}(t)$. Here, the trace is over medium-only states, $\hat{e}(t) = i\partial_t \hat{x}_{\0}^{}(t) - [\hat{x}_{\0}^{}(t),\hat{H}]$ is an error operator, and $\hat{\rho}_0= e^{-\beta\hat{H}_{0}}/Z_0$ is the medium-only density matrix with $Z_0=\text{Tr}[e^{-\beta\hat{H}_{0}}]$ the medium partition function in the grand canonical ensemble. By considering the stationary solutions, $\varphi_{0}^{}(t)= \varphi_{0}^{} e^{-i E t}$ and $\varphi_{\k\q}^{}(t) =  \varphi_{\k\q}^{}  e^{-i E t}$, we obtain the following eigenvalue problem:
\begin{subequations}
\label{eq:P3_finiteT}
\begin{align}
    E\varphi_{0}^{} &=  %
    -\Frac{v}{\area^2}\sum_{\k,\q} f_\q(1-f_\k) \varphi_{\k\q}^{}
\label{eq:P3_1} \\
    E\varphi_{\k\q}^{} &= E_{X\k\q} \varphi_{\k\q}^{} - v\varphi_{0}^{} %
    -\Frac{v}{\area}\sum_{\k'}(1-f_{\k'}) \varphi_{\k'\q}^{} %
    \; .
 \label{eq:P3_2}   
\end{align}
\end{subequations}
Here, $E_{X\k\q} = \epsilon_{X\q-\k} + \epsilon_\k - \epsilon_\q$ and we have used the Fermi-Dirac distribution for the electron occupation, $\langle \hat{c}^{\dag}_{\k} \hat{c}_{\k}^{} \rangle_{\beta}=f_\k = [e^{\beta (\epsilon_{\k}-\mu)}+1]^{-1}$, and  for the hole occupation, $\langle \hat{c}_{\k}^{} \hat{c}^{\dag}_{\k}\rangle_{\beta}=1-f_\k$. Note that we have dropped terms that vanish when $\Lambda \to \infty$; for instance, since $v\sim 1/\ln \Lambda \to 0$, terms like  
$\frac{v}{\area}\sum_{\q'} f_{\q'} \varphi_{\k\q'}^{}$ also go to zero as $\Lambda \to \infty$.

The set of equations in~\eqref{eq:P3_finiteT} constitutes an eigenvalue problem which can be solved to give a set of eigenvalues $E^{(n)}$ and associated eigenvectors $\varphi_{0}^{(n)}$ and $\varphi_{\k\q}^{(n)}$, with $n$ a discrete index. We require that the corresponding stationary operators
\begin{equation*}
    \hat{x}_{\0}^{(n)} = \varphi_{0}^{(n)} \hat{x}_{\0}^{} + \frac{1}{\area} \sum_{\k,\q} \varphi_{\k\q}^{(n)}  \hat{c}_{\q}^{\dag} \hat{c}_{\k}^{} \hat{x}_{\q-\k}^{}\; ,
\label{eq:stationary_imp-op}    
\end{equation*}
are orthonormal under a thermal average, $\langle \hat{x}_{\0}^{(n)} \hat{x}_{\0}^{(m)\dag} \rangle_{\beta} = \delta_{n,m}$, implying that
\begin{equation*}
    \varphi_{0}^{(n)} \varphi_{0}^{(m)*} + \frac{1}{\area^2}\sum_{\textbf{k},\textbf{q}} f_\q (1-f_\k) \varphi_{\k\q}^{(n)} \varphi_{\k\q}^{(m)*} = \delta_{n,m} \; .
\end{equation*}
The stationary operators thus form a complete basis within which we can expand the approximate impurity operator~\eqref{eq:var_imp-op}, giving
\begin{equation}
    \hat{x}_{\0}^{}(t) = \sum_n \varphi_{0}^{(n)*} \hat{x}_{\0}^{(n)} e^{-i E^{(n)} t}\; ,
\label{eq:Xexpansion}    
\end{equation}
where $\varphi_{0}^{(n)*} = \langle \hat{x}_{\0}^{} \hat{x}_{\0}^{(n)\dag}\rangle_{\beta}$ and where we have used the  boundary condition $\hat{x}_{\0}^{}(0) = \hat{x}_{\0}^{}$.

The exciton retarded Green's function in the time domain, 
$\mathcal{G}_{X} (t) = -i \theta(t) \langle [\hat{x}_{\0}^{}(t), \hat{x}_{\0}^{\dag}] \rangle_{\beta}=-i \theta(t) \langle \hat{x}_{\0}^{}(t) \hat{x}_{\0}^{\dag} \rangle_{\beta},$ can be evaluated approximately within the variational ansatz~\eqref{eq:var_imp-op} by using Eq.~\eqref{eq:Xexpansion}. By taking the Fourier transform into the frequency domain we obtain:
\begin{equation}
  G_{X} (\omega+i0) = \sum_n
  \frac{|\varphi_0^{(n)}|^2}{\omega-E^{(n)}+i0} \; ,
\end{equation}
where the small imaginary part originates from the Heaviside function $ \theta(t)$ in the retarded Green's function $\mathcal{G}_{X} (t)$. 

\subsection{Exciton self-energy and $T$ matrix}
\label{sec:T-matrix}
As discussed above, solving the eigenvalue problem~\eqref{eq:P3_finiteT} allows us to evaluate the exciton Green's function. It turns out that it is numerically convenient to instead consider the exciton self-energy $\Sigma(\omega)$, which is related to the Green's function via
\begin{equation}
    G_{X}(\omega) =\frac{1}{\omega - \Sigma(\omega)}\;.
\label{eq:dressedX}
\end{equation}
The expression for the exciton self-energy can be derived by manipulating the eigenvalue problem~\eqref{eq:P3_finiteT}~\cite{Liu2019}, following the same procedure valid at zero temperature~\cite{Combescot2007}  --- for completeness we describe this in Appendix~\ref{app:Xself-en}. In this way, the exciton self-energy reads
\begin{equation}
    \Sigma(\omega)=\frac{1}{\area}\sum_{\q} f_\q \mathcal{T}(\q, \omega+\epsilon_{\q})\; ,
\label{eq:Xself-energy}
\end{equation}
where the inverse of the $T$ matrix is defined as
\begin{equation}
    \mathcal{T}^{-1}_{}(\q, \omega) =
    -\frac{1}{v} - \frac{1}{\area} \sum_{\k} \frac{1-f_\k}{\omega - \epsilon_{\k} - \epsilon_{X\k-\q} + i0}\; .
\label{eq:invT-matrix}    
\end{equation}
The same expression~\eqref{eq:Xself-energy} can also be derived by using a diagrammatic expansion within the ladder approximation~\cite{Combescot2007,Liu2019}. Thus, our variational approach provides an additional theoretical foundation for the ladder diagrams.

It is profitable to separate the vacuum contribution to the $T$ matrix, describing the electron-exciton scattering in the absence of a surrounding Fermi gas, from the many-body contribution. To this end, we note that the logarithmic divergence of the second term in~\eqref{eq:invT-matrix} cancels with that of the inverse contact interaction constant $v^{-1}$~\eqref{eq:trion_binding-en}, allowing the vacuum contribution $\mathcal{T}_0$ to be calculated analytically~\cite{Levinsen2Dreview}:
\begin{subequations}
\label{eq:Tmatrix}
\begin{align}
   \mathcal{T}^{-1}_{}(\q, \omega) &=  \mathcal{T}^{-1}_{0}(\q, \omega)-\Pi_{mb}(\q, \omega) \\
   \mathcal{T}^{-1}_{0}(\q, \omega) &= \frac{m_r}{2\pi} \ln \left(\frac{-\varepsilon_T}{\omega - \frac{\q^2}{2m_T}+i0}\right)
   \label{eq:2b_Tmatrix}
   \\
\Pi_{mb}(\q, \omega) &= -\frac{1}{\area} \sum_{\k} \frac{f_\k}{\omega - \epsilon_{\k} - \epsilon_{X\k-\q} + i0} \label{eq:mb_Tmatrix}
   \; ,
\end{align}
\end{subequations}
where $m_T=m+m_X$ is the trion mass and $m_r=mm_X/m_T$ is the exciton-electron reduced mass. Note that the vacuum $T$ matrix~\eqref{eq:2b_Tmatrix} has a pole at the bound state, i.e., the trion energy, $\omega
=-\varepsilon_T+\epsilon_{T\q}$, where $\epsilon_{T\q} = \frac{q^2}{2m_T}$ is the trion kinetic energy (for a summary of the trion properties at finite momentum and finite doping, see Appendix~\ref{app:trion}). The many-body contribution 
$\Pi_{mb}(\q, \omega)$
to the $T$ matrix has %
to be evaluated numerically---for details on the numerical procedure see Appendix~\ref{app:mb_Tmatrix}.

\subsection{Absorption and photoluminescence}
\label{sec:spectrl-f}
The optical absorption coincides with
the exciton spectral function (up to a frequency-independent prefactor):
\begin{equation}
    A(\omega)=-\frac{1}{\pi}\Im G_X(\omega+i0) \; .
\label{eq:X_Spectr_func}
\end{equation}
Indeed, in the linear-response regime, %
the spectral function is equivalent to %
the transfer rate from an initial state $|n\rangle$ containing no excitons (the impurity vacuum) 
to a final state $|\nu\rangle$ containing a single exciton.
Here, the impurity vacuum and single-impurity states are eigenstates of the Hamiltonian, i.e., $\hat{H} |n\rangle =\hat{H}_{0} |n\rangle = E_n |n\rangle$ and $\hat{H} |\nu\rangle = E_\nu |\nu\rangle$.
Using Fermi's golden rule, we have
\begin{equation}
    A(\omega) = \sum_{n,\nu} \langle n| \hat{\rho}_{0} |n \rangle |\langle \nu| \hat{x}_{\0}^\dag |n \rangle |^2 \delta (E_{\nu} - E_n - \omega) \; ,
\label{eq:FermisGR_A}    
\end{equation}
where $\hat{\rho}_0$ is the medium-only partition function introduced above.
By using %
the completeness of the $\{|\nu\rangle \}$ basis, one can easily show that Eq.~\eqref{eq:FermisGR_A} coincides with~\eqref{eq:X_Spectr_func} 
--- see Ref.~\cite{Weizhe_PRA2020}. Using the  definition~\eqref{eq:FermisGR_A}, it is now straightforward to show that the exciton optical absorption satisfies the sum-rule:
\begin{equation}
    \int_{-\infty}^{\infty}d\omega\, A(\omega)=1\;.
\end{equation}

In order to evaluate the 
photoluminescence, 
we instead consider the opposite situation, i.e., an initial state $|\nu\rangle$ containing the medium and the exciton, and a final state $|n\rangle$ after the exciton has recombined to emit a photon.
Here, we have assumed that the exciton density is low enough such that  each exciton can be treated individually. The transfer rate at thermal equilibrium is then given by
\begin{equation}
    P(\omega) = \sum_{n,\nu} \langle \nu| \hat{\rho} |\nu \rangle |\langle n| \hat{x}_{\0}^{} |\nu \rangle |^2 \delta (E_{\nu} - E_n - \omega) \; ,
\label{eq:FermisGR_PL}    
\end{equation}
where $\hat{\rho} = e^{-\beta \hat{H}}/Z_{int}$ is the density matrix associated with the interacting exciton and medium system~\eqref{eq:Hamiltonian}  and $Z_{int} = \sum_\nu\expval*{e^{-\beta \hat{H}}}{\nu}$ the associated partition function. 
It is straightforward to show that the photoluminescence satisfies the following sum rule
\begin{equation}
    \int_{-\infty}^{\infty}d\omega \,P(\omega) = \text{Tr} [\hat{\rho} \hat{x}_{\0}^{\dag} \hat{x}_{\0}^{}] \; .
\end{equation}

Using the properties of the delta function, the absorption $A(\omega)$ and photoluminscence $P(\omega)$ can be related by a detailed balanced condition~\cite{Weizhe_PRA2020}:
\begin{equation}
    P(\omega) = \Frac{Z_0}{Z_{int}} e^{-\beta\omega} A(\omega) \; .
\label{eq:PL_func}
\end{equation}
The thermodynamic, Boltzmann-type scaling between absorption and emission profiles~\eqref{eq:PL_func} is also known as the  Kubo-Martin-Schwinger relation~\cite{Kubo_57,Martin-Schwinger_59}, the Kennard-Stepanov relation \cite{Kennard_18,Stepanov_57,McCumber_64} or the van Roosbroeck-Shockley
relation~\cite{vanRoosbroeck_54}, depending on the context within which it has been studied, and it applies to a broad range of systems, including semiconductors~\cite{Band-Heller_88,Chatterjee_PRL2004,Ihara_PRB2009}.
It relies on the assumption that the population of excited states, here excitons, has thermalized at a temperature $T$ before the emission and that they are otherwise uncorrelated. 
Note also that the thermalization temperature $T$ can be different from the system lattice (cryostat) temperature.

\subsection{Numerical implementation}
Even though one only has to evaluate two momentum integrals to obtain the exciton Green's function in Eq.~\eqref{eq:dressedX}, namely the integrals in Eqs.~\eqref{eq:Xself-energy} and \eqref{eq:mb_Tmatrix}, some comments about the numerical procedure are necessary. 
For the optical absorption~\eqref{eq:X_Spectr_func}, the numerical convergence of the integrals is much improved by shifting the frequency to the complex plane, $\omega \mapsto \omega + i\eta$. Apart from helping with convergence, this shift provides a simplified description of the exciton's intrinsic broadening due to effects beyond those included in the Hamiltonian such as recombination and disorder. In the following, we have used the typical value $\eta=0.04\varepsilon_T \simeq 1$~meV. 

Including $\eta$ implies that the exciton spectral function decays as a Lorentzian at low and high energies. However, one cannot evaluate the photoluminescence using this procedure. Indeed, by using the detailed balance condition~\eqref{eq:PL_func}, the photoluminescence diverges at infinitely low frequencies if one uses a finite value of $\eta$ to evaluate the absorption, since the Boltzmann occupation increases more rapidly than the Lorentzian decay of absorption. This means that photoluminescence needs to be evaluated by first calculating absorption at $\eta=0$ and then multiplying by the Boltzmann occupation. The effects of the exciton intrinsic decay time can be re-introduced at the end of the calculation by convolving the photoluminescence with a Lorentzian profile with broadening $2\eta$. This procedure is described in detail in  Appendix~\ref{app:mb_Tmatrix}. 

\begin{figure*}
    \centering
    \includegraphics[width=0.95\textwidth]{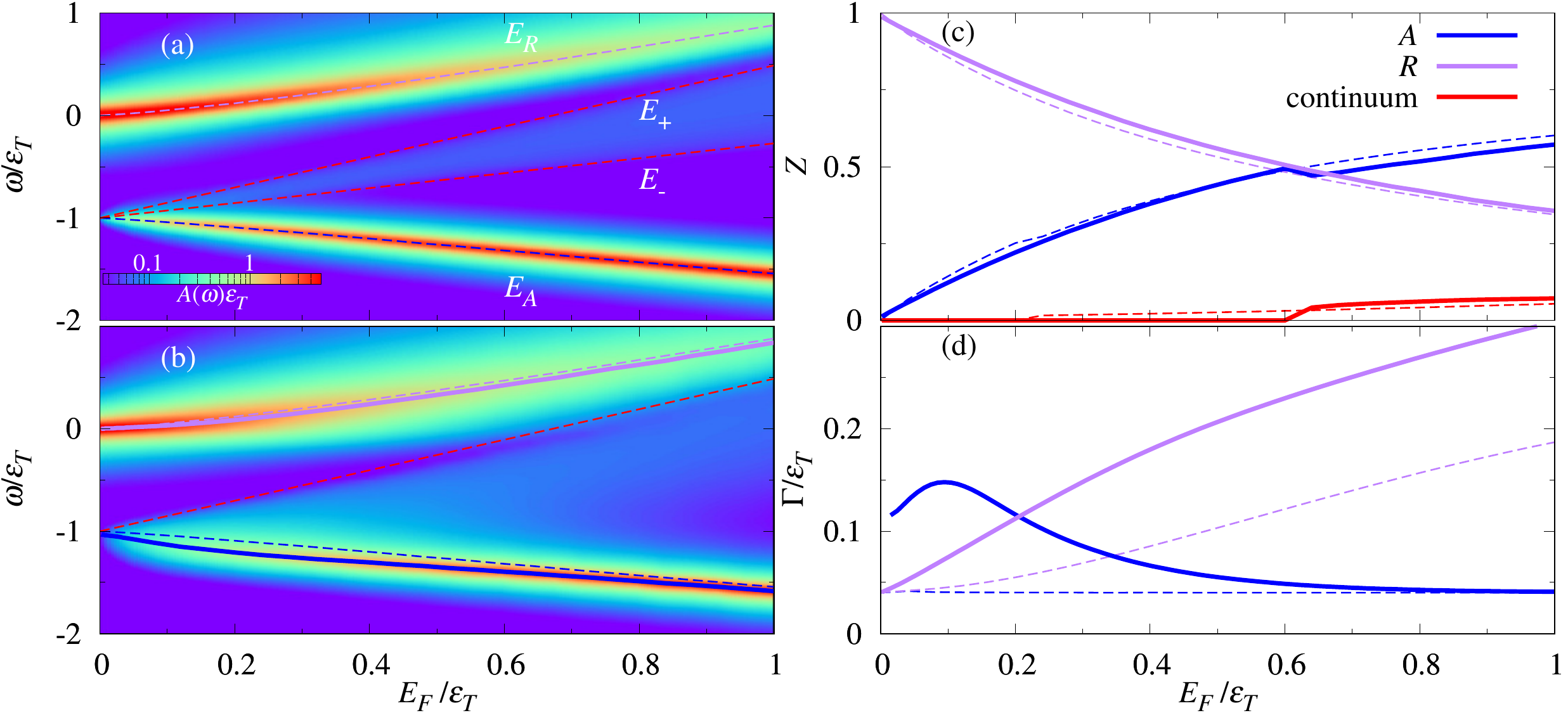}
    \caption{(a) Spectral function $A(\omega)$ at zero temperature as a function of doping and energy. Blue and purple dashed lines are respectively the attractive ($A$) and repulsive ($R$) polaron energies, while the dashed red lines are the boundaries of the trion-hole continuum $E_{\pm}$ (see text). (b) Spectral function $A(\omega)$ at temperature $T=50$~K$\simeq 0.17 \varepsilon_T$. Dashed lines are the zero-temperature energies as in panel (a) --- for the trion-hole continuum we plot only the upper boundary $E_+$. Solid lines are the attractive (blue) and repulsive (purple) branch energies at finite temperature. (c,d) Doping dependence of the spectral weights $Z$ and half linewidths $\Gamma$ extracted from the spectral function (see Appendix~\ref{app:properties} for details of how these are determined) at $T=0$ (dashed) and $T=50$~K$\simeq 0.17 \varepsilon_T$ (solid). In panel (d), note that the constant value of $\Gamma_A$ at $T=0$ coincides with the intrinsic broadening.
    }
    \label{fig:T50_vs_T0}
\end{figure*}

\section{Results}
\label{sec:results}
We now illustrate our results for both optical absorption and photoluminescence. 
In order to stress the differences between the zero and finite temperature cases, let us first briefly  summarize what is known about polarons at zero temperature. 

\subsection{Zero temperature}
\label{sec:T-0}
Numerous theoretical works have so far analyzed the zero-temperature properties 
of an impurity interacting with a free Fermi gas, in the context of both semiconductor heterostructures~\cite{Suris2003correlation,Sidler_NatPhys_2017,Efimkin2017,Rana_PRB2020,Rana_PRB2021,Rana_PRL2021,Efimkin_PRB_2021} and ultracold atomic gases~\cite{Massignan_RPP2014,Scazza2022}. 
The main polaron properties at $T=0$ are summarized in Fig.~\ref{fig:T50_vs_T0}. In panel (a) we plot the doping and energy dependence of the spectral function. One observes that the 
optical absorption is dominated by two polaron quasiparticle resonances: the attractive polaron branch at lower energy $E_{A}$ and the repulsive polaron branch at higher energy $E_R$. The energies $E_{A,R}$ are evaluated from the positions of the absorption peaks in Fig.~\ref{fig:T50_vs_T0}. 
Here, both polaron branches are quasiparticles in the sense that $E_{A,R}$  
satisfy the condition~\cite{fetterbook}
\begin{equation}
    E_{A,R} = \Re \Sigma (E_{A,R}) \; ,
\label{eq:GX_poles}   
\end{equation}
which coincides with a pole of the exciton Green's function when 
the corresponding imaginary part of the self-energy $\Im \Sigma (E_{A,R})$ becomes small. 
Note that the repulsive branch eventually stops satisfying this condition when $E_F\gtrsim 3\varepsilon_T$~\cite{Ngampruetikorn2012}, and thus ceases to be a polaron quasiparticle.
This large doping regime is not analyzed in this work.

In the limit of vanishing doping, the attractive branch energy recovers the trion  energy $E_A \to -\varepsilon_T$, while the repulsive branch energy reduces to the exciton energy, $E_R \to 0$. (We remind the reader that we measure energies with respect to that of the exciton at rest). However, differently from the trion, whose energy 
blueshifts with doping due to its interactions with the surrounding electrons %
[see Ref.~\cite{Parish_PRA11} and Eq.~\eqref{eq:trion-EF_Q01}], the attractive polaron branch redshifts with doping.  At the same time, the repulsive polaron branch %
blueshifts with doping.  Thus, at zero temperature, as soon as doping is finite,  the attractive and repulsive polaron quasiparticles are no longer the trion and the exciton, respectively, even if they recover the corresponding energies at low doping. Furthermore, within the single particle-hole-excitation ansatz~\eqref{eq:var_imp-op}, the effective mass of the attractive polaron~\cite{Schmidt_PRA2012,Parish_PRA2013} does not evolve into the trion mass at low doping.

As far as the coupling to light of the  polaron branches is concerned, at very low doping the repulsive branch retains all the spectral weight and the attractive branch is dark, as is also the case for the trion. However, when doping increases, there is a transfer of oscillator strength from the repulsive to the attractive branch. 
This is shown in Fig.~\ref{fig:T50_vs_T0}(c), where we plot the doping dependence of the spectral weights $Z_{A,R}$ for each branch. At low doping $E_F \ll \varepsilon_T$, $Z_A$ grows 
linearly with $E_F$, which agrees with the doping dependence of the trion oscillator strength~\cite{GlazovJCP2020}.

There are also important differences between the attractive and repulsive polaron branches at zero temperature, even though they both satisfy  Eq.~\eqref{eq:GX_poles} for values of doping up to  $E_F \sim \varepsilon_T$. The  attractive polaron is always a sharp resonance, with a Lorentzian broadening coinciding with the intrinsic broadening $2\Gamma_A=2\eta$ --- see Fig.~\ref{fig:T50_vs_T0}(d). By solving the eigenvalue problem~\eqref{eq:P3_finiteT} at zero temperature, one finds that the attractive branch energy coincides with the lowest eigenvalue energy, $E_A=E^{(n=1)}$, which is separated from the energy of the excited states $n>1$ by a gap.
By contrast, the repulsive branch %
does not correspond to a specific %
eigenstate; rather it is composed of a continuum of eigenstates with closely spaced eigenvalues, resulting in a polaron quasiparticle with finite lifetime. As a consequence, its broadening $2\Gamma_R$ grows monotonically with $E_F$ and only coincides with $2\eta$ at small doping  --- see Fig.~\ref{fig:T50_vs_T0}(d).

In between the attractive and repulsive branches, one can observe in Fig.~\ref{fig:T50_vs_T0}(a) a continuum of trion and Fermi-sea-hole states with a well-defined relative momentum $\q$ in the variational function $\varphi_{\k\q}^{}$. The boundaries of this so-called trion-hole continuum can be evaluated from the energy of a trion (see Appendix~\ref{app:trion}) plus a hole separately. If the hole has zero momentum $\q=\0$, then, because of momentum conservation, the trion is also at zero %
momentum and the upper boundary of the trion-hole continuum is
\begin{equation}
    E_{+} = E_T^{(\0,E_F)}=-\varepsilon_T + \frac{m_T}{m_X} E_F\; .
\label{eq:Eplus}    
\end{equation}
Conversely, the energy of the trion-hole lower bound is
\begin{equation}
    E_{-} = E_T^{(\k_F,E_F)} -E_F\; ,
\label{eq:Eminus}    
\end{equation}
where both the hole and trion are at $\q=\k_F=k_F \hat{\textbf{n}}$, with $\hat{\textbf{n}}$ an arbitrary direction, and $E_T^{(\k_F,E_F)}$ is the trion energy [which can be evaluated from Eq.~\eqref{eq:trion_finiteQ-EF}]. Thus, at zero temperature and finite doping, the attractive branch is separated from  the trion-hole continuum by an energy gap. Note that, while this is a consequence of considering a single excitation of the medium, our results are consistent with those of diagrammatic quantum  Monte Carlo~\cite{Goulko_PRA2016} which demonstrated that there is a region of anomalously low spectral weight (a ``dark continuum'') above the narrow attractive polaron branch. Furthermore, calculations that incorporate Coulomb interactions between charges also obtain a suppression of spectral weight of this continuum~\cite{Rana_PRB2020}.

\begin{figure}
    \centering
    \includegraphics[width=1\columnwidth]{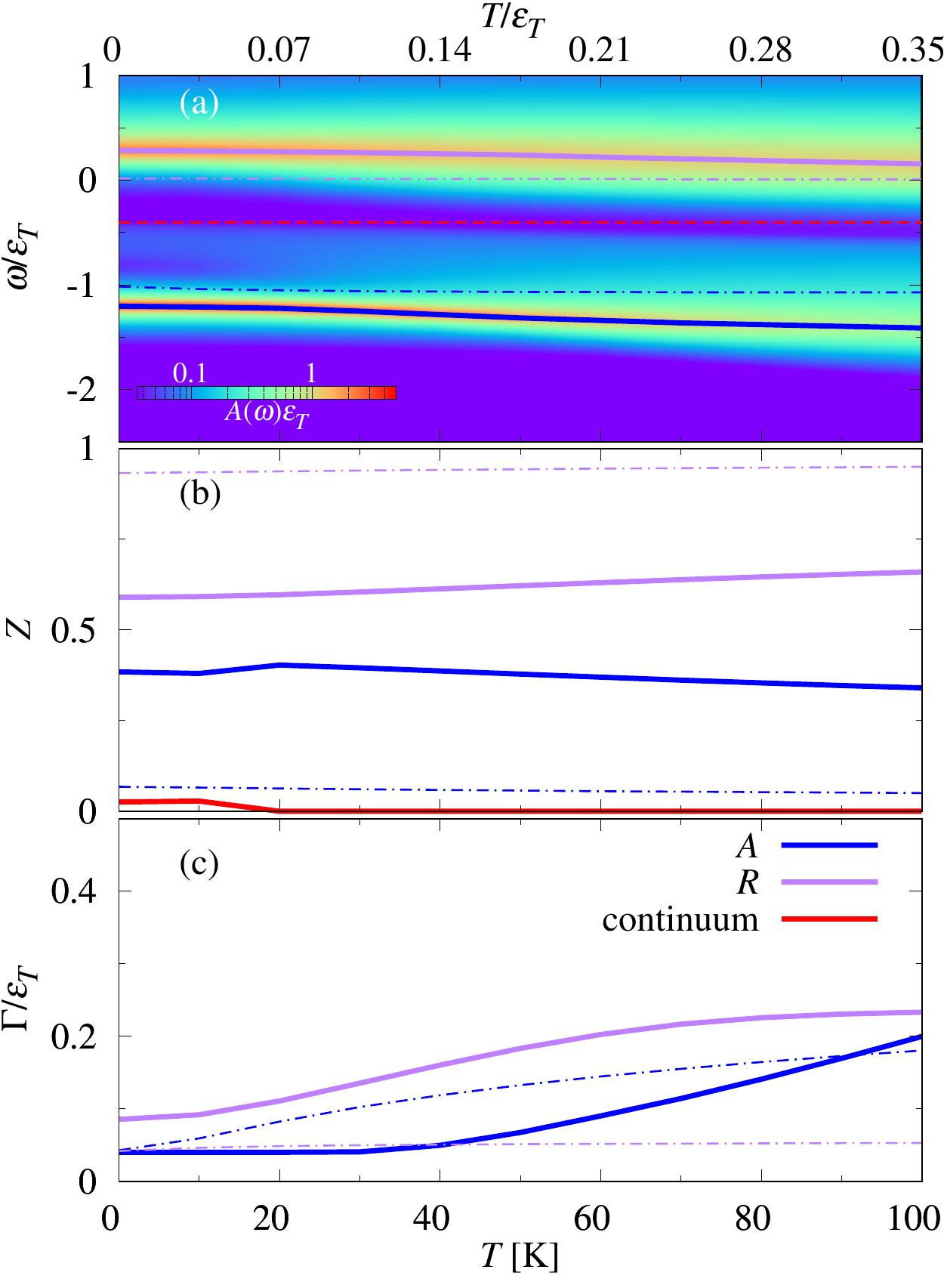}
    \caption{(a) Spectral function $A(\omega)$ at finite doping $E_F=0.4\varepsilon_T$ as a function of temperature and energy. The solid lines are the energies of the attractive ($A$, blue) and repulsive ($R$, purple) branches, while the dashed (red) line is the $T=0$ upper boundary of the trion-hole continuum $E_{+}$ in Eq.~\eqref{eq:Eplus}. Dot-dashed lines are the $A$ and $R$ energies at $E_F=0.04\varepsilon_T$. (b,c) Temperature dependence of spectral weight $Z$ and half linewidth $\Gamma$ evaluated from the exciton spectral function (see Appendix~\ref{app:properties}) at $E_F=0.04\varepsilon_T$ (dot-dashed) and $E_F=0.4\varepsilon_T$ (solid). Note that, in panel (c), the constant value of $\Gamma_R$ at $E_F=0.04\varepsilon_T$ is approximately given by the intrinsic broadening $\eta$.
    }
    \label{fig:EF0.4_Properties}
\end{figure}
\subsection{Finite temperature}
We now discuss the effect of temperature on the optical response.  
Figure~\ref{fig:T50_vs_T0}(b-d) allows a comparison between the doping-dependent properties of optical absorption at zero and finite temperature.
Both the energies and spectral weights of attractive and repulsive branches have a very weak dependence on temperature in the regime $T\lesssim \varepsilon_T$. The branch energies are slightly redshifted, while $Z_A$ ($Z_R$) is slightly smaller (larger) compared to the zero-temperature case. This small variation with temperature is also observed in Fig.~\ref{fig:EF0.4_Properties}(a,b) for a fixed doping.
The most important difference  at finite temperature is the behavior of the trion-hole continuum, which  %
subsumes the attractive branch when $T\gtrsim E_F$, i.e., at sufficiently low doping or sufficiently high temperature. 
This can be clearly seen from Fig.~\ref{fig:T50_vs_T0}(b), where we no longer observe a sharp lower bound %
of the  
trion-hole continuum  at low doping because, at finite temperature, the unbound hole belonging to the trion-hole continuum can thermally occupy any momentum state. However, the upper bound of the trion-hole continuum is still clearly visible and approximately follows  
the zero-temperature expression  $E_{+}$~\eqref{eq:Eplus}. Similarly, for a fixed doping in Fig.~\ref{fig:EF0.4_Properties}(a), we observe that the trion-hole continuum is only well-separated from the attractive branch at low temperatures.

Since the spectral weight of the trion-hole continuum is small, its merging with the  attractive branch only slightly affects the attractive peak energy. 
However, the disappearance of the attractive polaron quasiparticle strongly modifies the attractive-branch linewidth $2\Gamma_A$.  %
In particular, we observe in Fig.~\ref{fig:T50_vs_T0}(d) that $\Gamma_A$ has a striking  non-monotonic dependence at low doping, while it decreases towards its zero-temperature value (corresponding to the intrinsic broadening $\eta$) when $E_F$ increases. 
Likewise, increasing the temperature at fixed doping can substantially increase $\Gamma_A$ from $\eta$, as shown in Fig.~\ref{fig:EF0.4_Properties}(c).
As we will discuss in Sec.~\ref{sec:no-QP}, this behavior signals a crossover from a coherent Fermi polaron regime to an incoherent trion-dominated regime, 
where there no longer exists a well-defined attractive quasiparticle that is separated from the trion-hole continuum.

The repulsive branch, on the other hand, remains a polaron quasiparticle with a finite lifetime (broadening) %
for the dopings considered in this work ($E_F \lesssim \varepsilon_T$). In particular, we see that temperature does not change the nature of the repulsive branch in Figs.~\ref{fig:T50_vs_T0} and \ref{fig:EF0.4_Properties}, but it can lead to a faster increase of the half linewidth $\Gamma_R$ with increasing doping [Fig.~\ref{fig:T50_vs_T0}(d)].

\begin{figure}
    \centering
    \includegraphics[width=1.0\columnwidth]{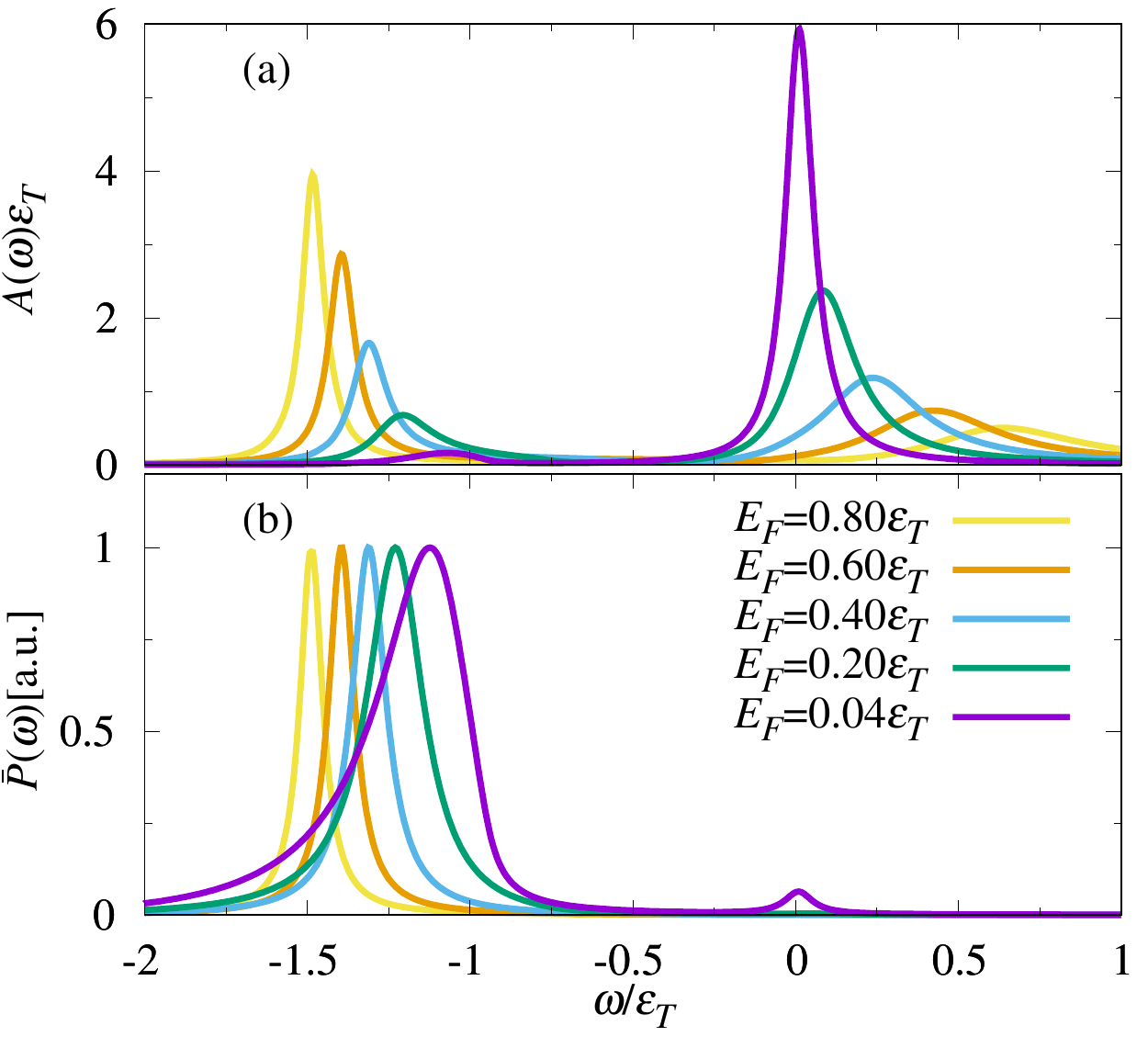}
    \caption{(a) Spectral function $A(\omega)$ and (b) Lorentzian convolved photoluminescence $\bar{P}(\omega)$
    for different dopings $E_F$ and at a fixed temperature of $T=50$~K$\simeq 0.17 \varepsilon_T$. The attractive branch photoluminescence peaks are rescaled to unity. 
    }
    \label{fig:T50_Spectr}
\end{figure}

\begin{figure}
    \centering
    \includegraphics[width=1.0\columnwidth]{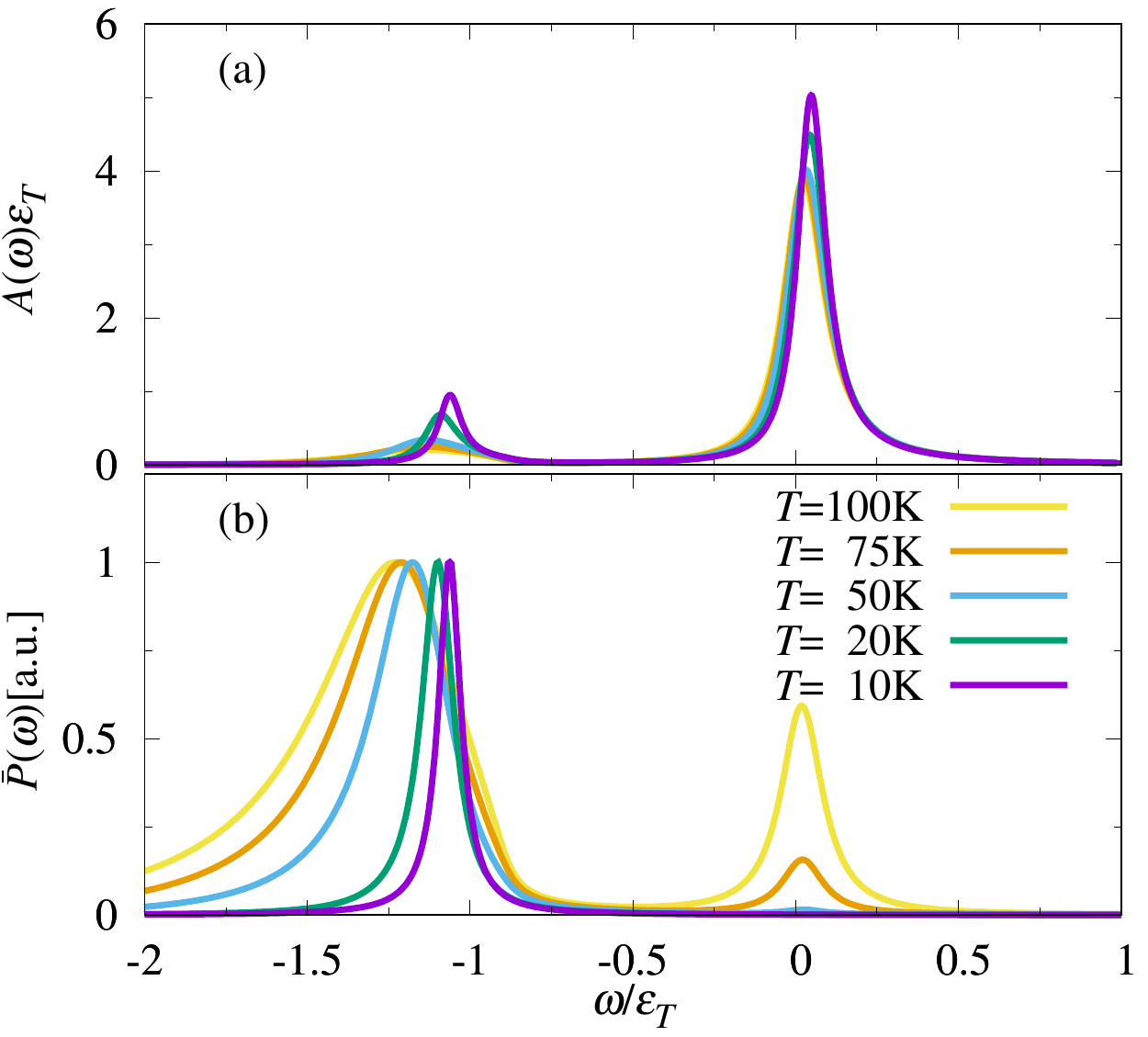}
    \caption{
    (a) Spectral function $A(\omega)$ and (b) Lorentzian convolved photoluminescence $\bar{P}(\omega)$
    for different values of the temperature $T$ and at a fixed doping $E_F=0.1\varepsilon_T$. The attractive branch photoluminescence peaks are rescaled to unity. 
    }
    \label{fig:EF0.04_Spectr}
\end{figure}

\begin{figure}
    \centering
    \includegraphics[width=1.0\columnwidth]{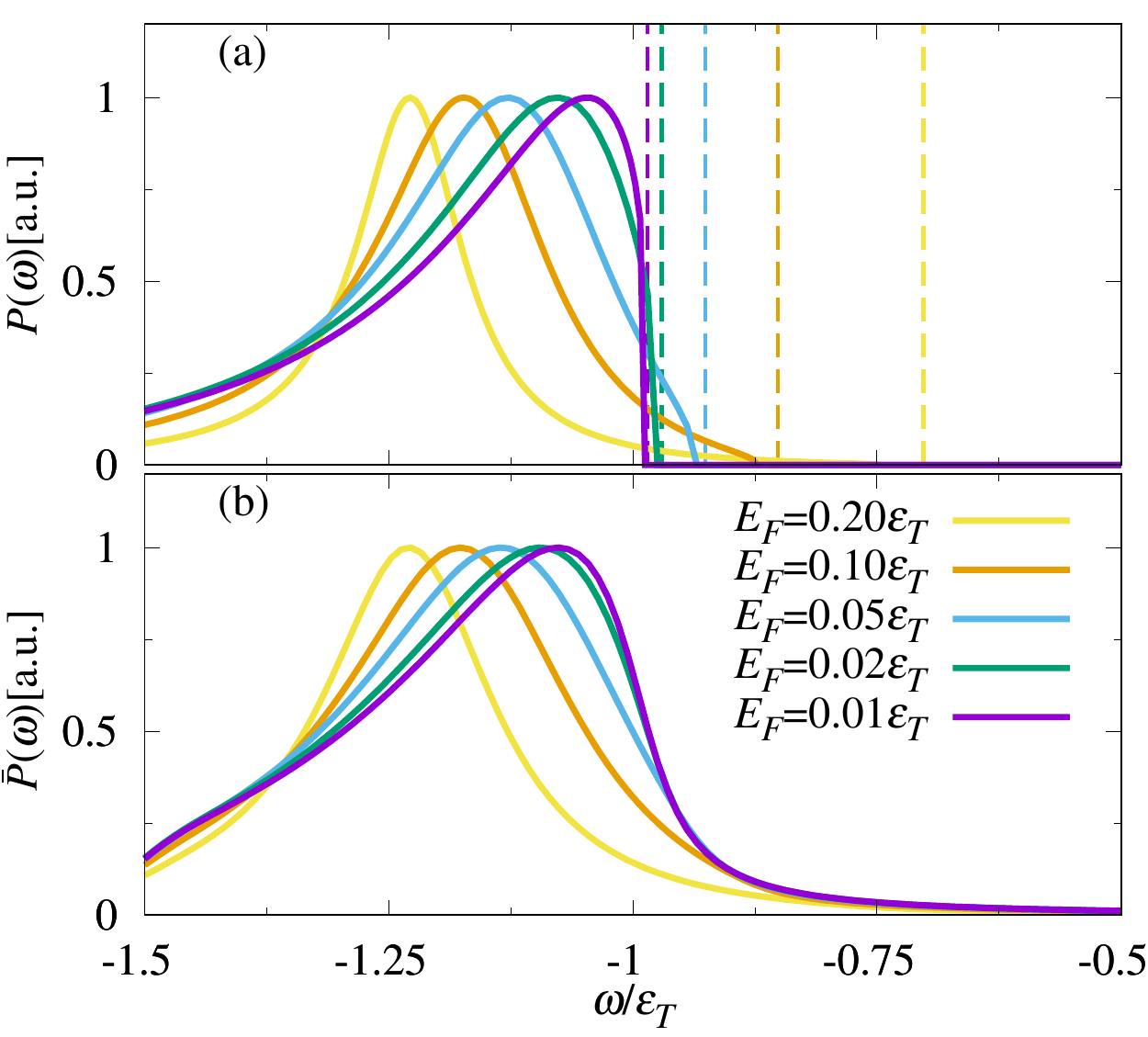}
    \caption{(a) Unconvolved photoluminescence $P(\omega)$ and (b) Lorentzian convolved photoluminescence $\bar{P}(\omega)$ for different dopings $E_F$, at a fixed temperature of $T=50$~K$\simeq 0.17 \varepsilon_T$, and for a frequency range around the attractive branch only. The photoluminescence peaks are rescaled to unity. In panel (a) the vertical dashed lines are the upper boundary of the trion-hole continuum $E_{+}$ at zero temperature (see text). 
    }
    \label{fig:T50_Offset}
\end{figure}
We now discuss the shape of %
the optical response %
profiles and how they evolve  
with either doping or temperature.
Figures~\ref{fig:T50_Spectr} and \ref{fig:EF0.04_Spectr} display both the absorption and the Lorentzian convolved photoluminescence (see Appendix~\ref{app:mb_Tmatrix}) %
at fixed temperature and fixed doping, respectively. 
The repulsive polaron quasiparticle shows up approximately as a Lorentzian symmetric profile  in both absorption and photoluminescence spectra, with a full width at half maximum (FWHM) that increases with increasing $E_F$. %
Note that the FWHM of the repulsive branch only has a weak dependence on temperature in Fig.~\ref{fig:EF0.04_Spectr} since $E_F$ has been fixed to a low value such that the intrinsic width $2\eta$ dominates [see  Fig.~\ref{fig:EF0.4_Properties}(c)]. 
By contrast, the shape of the attractive branch is strongly modified by temperature: in the low-temperature (high-doping) regime $E_F > T$, it is described by a Lorentzian with FWHM $2\eta$, while for $E_F < T$, it develops a strongly asymmetric shape with an exponential tail below the trion energy.
This evolution in the asymmetry of the attractive branch once again indicates a crossover from a Fermi-polaron quasiparticle to a continuum of trion states.

In Fig.~\ref{fig:T50_Offset} we further analyze the shape of the attractive peak at low doping $E_F \lesssim T$, 
comparing the Lorentzian convolved photoluminescence $\bar{P} (\omega)$ with %
the ``bare'' %
photoluminescence $P(\omega)$, where we have removed the effects of any intrinsic exciton  %
broadening. In panel (a), we observe a sharp onset of the photoluminescence which approximately coincides with the upper boundary of the trion-hole continuum at zero temperature, $E_+$. Thus, according to Eq.~\eqref{eq:Eplus}, it blueshifts with increasing doping. %
As shown in Fig.~\ref{fig:T50_Offset}(b), any intrinsic broadening $\eta$ 
only 
smooths out the sharp onset, while it has little effect on the position of the peak.
When $E_F$ increases, the sharp onset tends to disappear as the attractive peak redshifts and detaches from the trion-hole continuum.

Our calculated profiles for the attractive branch are in excellent quantitative agreement with recent experiments in the high-temperature (low-doping) regime~\cite{Zipfel_PRB2022}, as discussed in the accompanying paper~\cite{finiteTshort}. 
The exponential tail of the asymmetric attractive peak has previously been modelled within a trion picture for the case of  photoluminescence~\cite{Esser_PRB2000,Esser2001,Ross2014,Christopher2017}. There, the tail is ascribed to the kinetic energy of remaining electrons after the exciton within each trion has decayed into a photon. This description in terms of electron recoil can be formally derived from our theory in the limit of weak interactions, as we will discuss in Sec.~\ref{sec:virial}.

\subsection{Loss of the attractive polaron quasiparticle: polaron to  trion-hole continuum crossover}
\label{sec:no-QP}
\begin{figure}
    \centering
    \includegraphics[width=1.0\columnwidth]{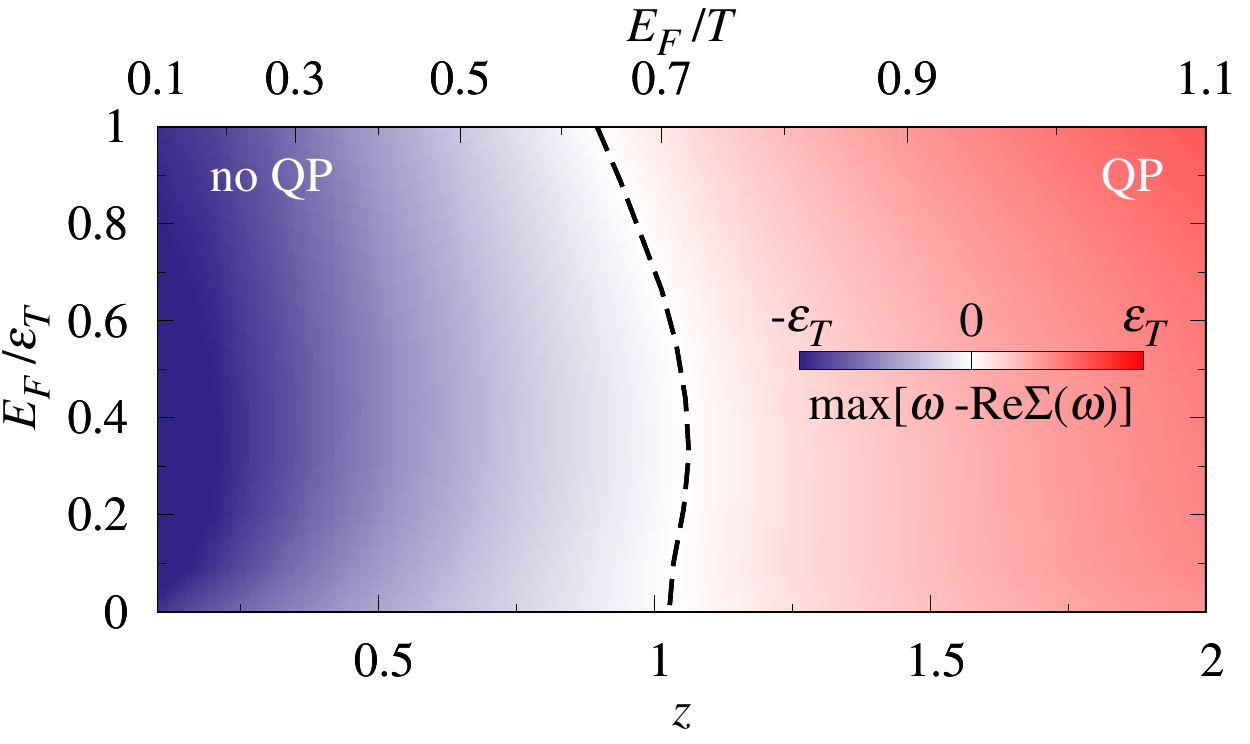}
    \caption{Maximum value of $\omega - \Re \Sigma(\omega)$ for $\omega < 0$ as a function of the Fermi gas fugacity $z=e^{\beta \mu}$ and doping. The black dashed line describes the values of $z$ and $E_F$ at which this maximum is zero. On the left of this curve (blue area) the attractive branch is not a polaron quasiparticle (no QP), %
    while on the right (red area) the attractive branch is a well defined polaron quasiparticle (QP).
    }
\label{fig:QP-noQP}
\end{figure}

In this section, we use the pole condition in Eq.~\eqref{eq:GX_poles} to characterize the crossover from a well-defined polaron quasiparticle to a trion-hole continuum with increasing temperature (decreasing doping).
In order to find the values of temperature and doping at which this crossover occurs, we plot in Fig.~\ref{fig:QP-noQP} the local maximum value of the function $\omega - \Re \Sigma (\omega)$ for $\omega < 0$ %
and identify the curve of doping versus fugacity $z=e^{\beta \mu} = e^{\beta E_F} -1$ at which this maximum value is zero. For $E_F \lesssim \varepsilon_T$, we find that this occurs roughly when $z \sim 1$ and thus $E_F \sim 0.7 T$. On the left of this curve, we lose the attractive polaron quasiparticle, as the condition~\eqref{eq:GX_poles} cannot be satisfied, i.e., $E_A - \Re \Sigma(E_A)\ne 0$. On the right of this curve, instead, the system is in the polaron regime where~\eqref{eq:GX_poles} is satisfied.

In order to further illustrate this crossover, we compare the results for the polaron energies, spectral weights and linewidths 
extracted from the spectral function with those obtained by treating the polaron as a well-defined quasiparticle~\cite{fetterbook}.  %
In the latter case, the polaron properties can be obtained directly from the expression of the impurity self-energy. Close to a quasiparticle %
resonance, the exciton Green's function can be approximated as
\begin{equation}
\label{eq:poleApprox}
    G_X(\omega) \Simiq_{\omega \simeq E_{j}} \frac{Z_{j}}{\omega-E_{j} +i\Gamma_{j}} \; ,
\end{equation}
where $j=A,R$ is the two-branch index, and the quasiparticle energy $E_{j}$ is a solution of  Eq.~\eqref{eq:GX_poles}. The pole weight or residue $Z_{j}$ is
\begin{equation}
    Z_{j} = \left(1 - \left.\Frac{\partial \Re  \Sigma(\omega)}{\partial \omega}\right|_{E_{j}}\right)^{-1} \; ,
\label{eq:FL_Z}    
\end{equation}
 and the polaron damping rate is
\begin{equation}
    \Gamma_{j} = - Z_{j}\Im \Sigma(E_{j})
    \; .
\label{eq:FL_Gamma}
\end{equation}

\begin{figure}
    \centering
    \includegraphics[width=1.0\columnwidth]{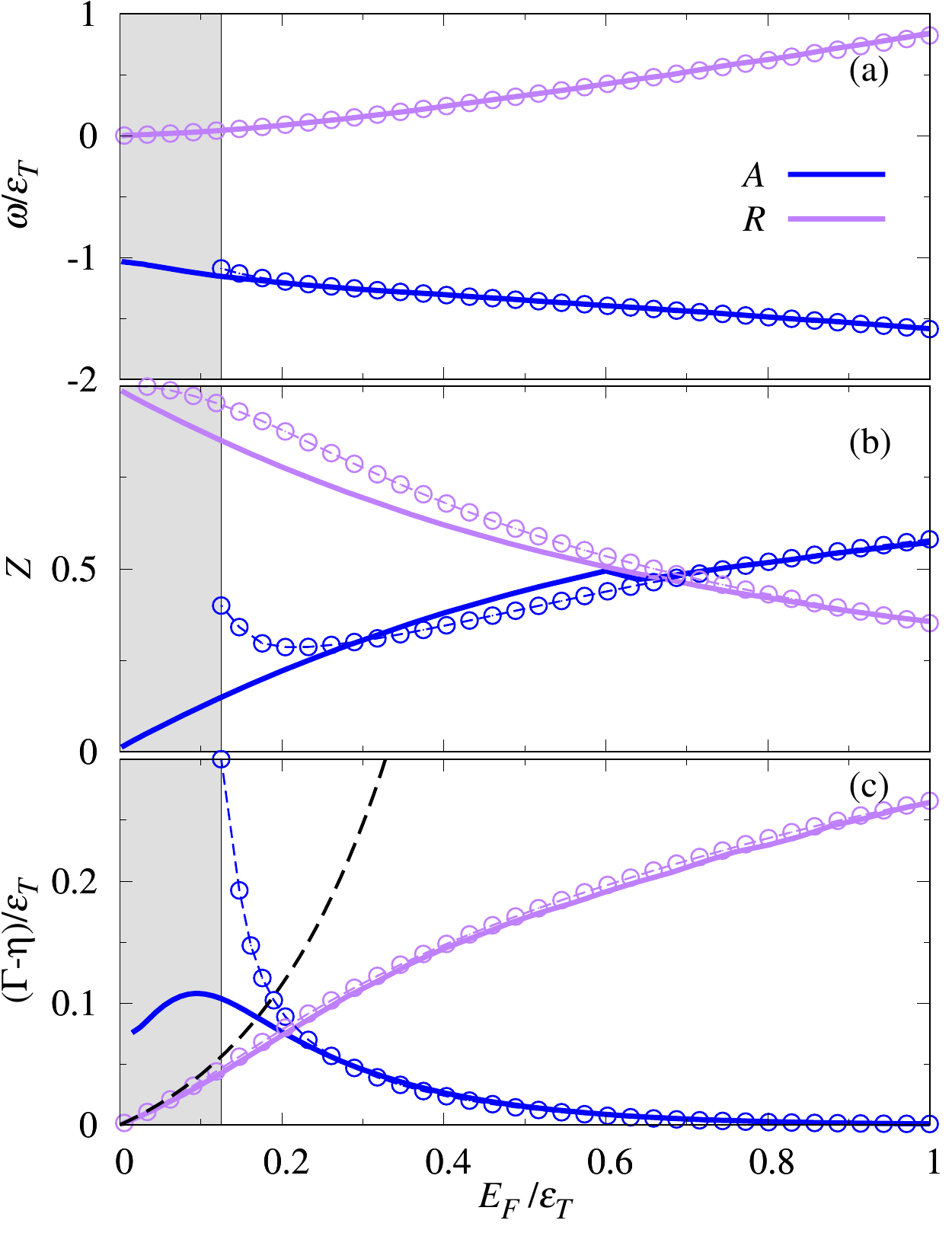}
    \caption{(a) Polaron energies $E_{A,R}$, (b) spectral weights $Z_{A,R}$, and (c) half linewidths $\Gamma_{A,R}$, evaluated from the exciton spectral function (solid lines) (as described in Appendix~\ref{app:properties}) and from %
    quasiparticle expressions (symbols). The attractive branch stops to be a quasiparticle resonance in the gray region at low doping when $E_F \lesssim 0.7 T$. 
    In panel (c) we plot $\Gamma_{A,R}-\eta$ is order to compare the numerical results with the analytical estimate of the repulsive branch broadening evaluated at $\eta=0$ at small doping (dashed line), derived within the virial expansion in Sec.~\ref{sec:virial}.
    Temperature is fixed at $T=50$~K$\simeq 0.17 \varepsilon_T$.}
\label{fig:compare_FL-th}
\end{figure}

We compare the results for $E_j$, $Z_j$, and $\Gamma_j$ obtained with both methods in Fig.~\ref{fig:compare_FL-th}. We observe that the positions of the poles coincide to high accuracy with those of the spectral function maxima 
--- see panel (a). 
For the repulsive branch, both the spectral weight and the broadening %
from the quasiparticle theory %
are in good agreement with those evaluated from the spectral function, even when the linewidth is non-negligible. By contrast, for the attractive branch, the results depart from one another when we approach the (gray) region $E_F \lesssim 0.7 T$ where there is no attractive quasiparticle, and the quasiparticle description breaks down since, according to Eq.~\eqref{eq:FL_Z}, $1/Z_A\to0$ when $\rm{max}[\omega-\Re\Sigma(\omega)]=0$. %

In the following section, we analyze the system properties well inside the trion-hole continuum (gray) regime, where the attractive branch is no longer a polaron quasiparticle. Here, for temperatures $T \gtrsim E_F$, we can apply a systematic quantum virial expansion. This connects the results of this work with those obtained in the accompanying paper~\cite{finiteTshort}.


\subsection{Virial expansion and connection to the trion wave function at high temperature or low doping}
\label{sec:virial}
As discussed, at high temperature or low doping such that the fugacity $z= e^{\beta \mu}\lesssim 1$, the attractive polaron quasiparticle disappears and the attractive branch only consists of a broad continuum. In this limit, one can formally apply a perturbatively exact quantum virial expansion in the fugacity~\cite{finiteTshort}. We will now briefly discuss how this expansion is related to the polaron theory, and how this allows us to demonstrate that the trion picture results of Refs.~\cite{Esser_PRB2000,Zipfel_PRB2022} are contained within the polaron formalism.

When $T \gg E_F$ and $z \simeq \beta E_F\ll 1$, we can formally expand the Fermi occupation function which then coincides with the Boltzmann distribution $f_\k\simeq z e^{-\beta\epsilon_{\k}}$. Likewise, to leading order in $z$ we have ${\mathcal T}\simeq {\mathcal T}_0$ --- see Eq.~\eqref{eq:Tmatrix}. Within this expansion, 
the exciton self-energy in Eq.~\eqref{eq:Xself-energy} becomes
\begin{align}
    \Sigma(\omega)\simeq \frac z\area\sum_\q e^{-\beta\epsilon_\q} {\mathcal T}_0(\q,\omega+\epsilon_\q)\;.
    \label{eq:SEsmallz}
\end{align}
In other words, to leading order in the fugacity, the self-energy is determined by two-body interactions weighted by a Boltzmann distribution.

Focusing first on the attractive branch, the dominant contribution to the self-energy arises from the pole of the $T$ matrix at the trion energy. We therefore expand the $T$ matrix for $\omega \simeq -\varepsilon_T + \epsilon_{T\q}$, with the result
\begin{align}\label{eq:Tvacpole}
    \mathcal{T}_{0}(\mathbf{q},\omega+\epsilon_\q) \simeq%
    \frac{Z_T}{\omega+\epsilon_{\q}-\epsilon_{T\q}+\varepsilon_T+i0}\;,
\end{align}
where $Z_T\equiv 2\pi \varepsilon_T/m_r$. %
The $T$ matrix can be expressed in terms of the vacuum ($E_F=0$) trion wave function $\tilde{\eta}_{\q_r}^{(\0)}$ for a contact electron-exciton interaction (see Appendix~\ref{app:Tfromwave}).
In our case, the relative momentum is $\q_r=\q m_X/m_T$, and therefore the kinetic energy of the relative motion is $\epsilon_{r\q_r}=\epsilon_\q-\epsilon_{T\q}$. Using the expression for the vacuum trion wave function %
in the center of mass frame, $\tilde{\eta}_{\q_r}^{(\0)}= \Frac{\sqrt{Z_T}}{\varepsilon_T + \epsilon_{r\q_r}}$ (see Appendix~\ref{app:trion}), we have
\begin{align}
    \mathcal{T}_0(\mathbf{q},\omega+\epsilon_\q) \simeq \frac{|(\varepsilon_T+\epsilon_{r\q_r})\tilde{\eta}_{\q_r}^{(\0)}|^2}{\omega+\epsilon_{r\q_r}+\varepsilon_T+i0}\;.
    \label{eq:Tvacwave}
\end{align}
Here, the numerator is derived by manipulating the relation between the trion wave function and $Z_T$, and is momentum independent. However, this is a special property of the contact electron-exciton interaction, %
and Eq.~\eqref{eq:Tvacwave} in fact yields the correct generalization for an arbitrary electron-exciton interaction that leads to the formation of a trion. In other words, it can be applied for any realistic trion wave function. %
For details of the generalization to arbitrary interactions, see Appendix~\ref{app:Tfromwave}.

Using the approximation for the vacuum $T$ matrix, Eq.~\eqref{eq:Tvacwave}, the self-energy in Eq.~\eqref{eq:SEsmallz} becomes
\begin{multline}
    \Sigma_{A}(\omega<-\varepsilon_T) \simeq
    \frac z\area\sum_\q e^{-\beta\epsilon_\q} |\tilde{\eta}_{\q_r}^{(\0)}|^2  \\
    \times
    \left[{\mathcal P}\frac{|\epsilon_{r\q_r}+\varepsilon_T|^2}{\omega+\epsilon_{r\q_r}+\varepsilon_T}-i\pi\omega^2 \delta(\omega+\epsilon_{r\q_r}+\varepsilon_T)\right]\;,
    \label{eq:SigmaTrionWF}
\end{multline}
where ${\mathcal P}$ denotes the principal value. %
This explicitly relates the self-energy calculated within the polaron theory to the trion wave function in the regime where the fugacity is small, and importantly it applies for any realistic trion wave function. A similar approach involving trion wave functions has been used to calculate absorption~\cite{Bronold2000}. Note that the real part diverges logarithmically when $\omega+\varepsilon_T\to0^-$, and therefore it cannot in general be neglected close to the onset of the attractive branch. 

For the repulsive branch, we can again apply Eq.~\eqref{eq:SEsmallz} to find the leading contribution at small fugacity. In the regime where $E_F\ll \varepsilon_T$, 
the width of the repulsive branch is much smaller than the trion binding energy, and to lowest order in the fugacity, we can simply evaluate the repulsive polaron self-energy within the repulsive branch by taking $\omega=0$. We then use the fact that, to logarithmic accuracy, the logarithmic behavior at small momentum is generic for the vacuum $T$ matrix for any short-range interaction%
~\cite{AdhikariAJP86,landau2013quantum} and thus we have %
\begin{align}
    \Sigma_{R}(0) & \simeq \frac z\area\frac{2\pi}{m_r}\sum_\q \frac{e^{-\beta\epsilon_\q}}{\ln(\varepsilon_T/\epsilon_{r\q_r})+i\pi} \nn \\ 
    & \simeq \frac{z(m/m_r) T}{\pi^2+\ln^2(e^{\gamma_{\rm E}}\beta\varepsilon_T)}\left[\ln(e^{\gamma_{\rm E}}\beta \varepsilon_T)-i\pi\right]\; ,
\label{eq:virialrep}
\end{align}
where $\gamma_{\rm E}\simeq 0.5772$ is the Euler-Mascheroni constant. Here, we have evaluated the integral by noting that  the integral is dominated by $\epsilon_\q\sim 1/\beta$ for small $\beta\varepsilon_T$, and the inclusion of $\gamma_{\rm E}$ originates from expanding the integral to the first subleading order in $\beta\varepsilon_T$.

The self-energies in Eqs.~\eqref{eq:SigmaTrionWF} and \eqref{eq:virialrep} can now be directly inserted into the Dyson equation~\eqref{eq:dressedX} to yield the absorption %
in Eq.~\eqref{eq:X_Spectr_func} or the photoluminescence in Eq.~\eqref{eq:PL_func}. To be explicit, within the virial expansion we have the exciton spectral function
\begin{align}
    A(\omega) = -\frac1\pi {\rm Im} \left[\frac{\Theta(-\omega-\varepsilon_T)}{\omega-\Sigma_{A}(\omega)}+\frac1{\omega-\Sigma_{R}(0)}\right].
\end{align}
This yields a broad continuum for the attractive branch, where the spectral weight vanishes at  $\omega+\varepsilon_T\to0^-$, and %
when we are far below this onset, 
we have $\Sigma_{A}(\omega)/\omega\to0$. Therefore we have an exponential tail modulated by the trion wave function:
\begin{align}
    A(\omega)\Simiq_{\omega \to -\infty} z\, e^{\beta (\omega +\varepsilon_T)\frac{m_T}{m_X}}\left|\tilde{\eta}_{\sqrt{2m_r|\omega+\varepsilon_T|}}^{(\0)}\right|^2.
\end{align}
The peak of the absorption is between the onset and the tail, and therefore it will not correspond to the vacuum trion energy. This can lead to an overestimate of the trion binding energy in experiments~\cite{Zipfel_PRB2022}, as shown in the accompanying paper~\cite{finiteTshort}. 
From Eq.~\eqref{eq:virialrep}, we find that the repulsive branch is a Lorentzian of width $\Gamma_{R}=\pi(m/m_r) \ef/[\pi^2+\ln^2(e^{\gamma_{\rm E}}\beta\varepsilon_T)]$. 
We have compared in Fig.~\ref{fig:compare_FL-th}(c) this expression against the numerical evaluation of the repulsive branch half linewidth $\Gamma_R - \eta$ (where we have removed the effect of the intrinsic exciton lifetime), finding excellent agreement at low doping, inside the region of validity of the virial expansion.

For the photoluminescence, we find
\begin{align}
    P(\omega) \simeq -\frac{Z_0}{Z_{int}}\frac1\pi {\rm Im} \left[\frac{\Theta(-\omega-\varepsilon_T)e^{-\beta\omega}}{\omega-\Sigma_{A}(\omega)}+\frac1{\omega-\Sigma_{R}(0)}\right],
\end{align}
where we have used the fact that the width of the repulsive branch is much smaller than the temperature, and thus the repulsive branch is very weakly modified by the Boltzmann factor.

\subsection{Connection to the trion theory of electron recoil}
Finally, we discuss how our theory relates to the calculation of electron recoil in previous trion-picture calculations such as by Esser \textit{et al.}, Refs.~\cite{Esser_PRB2000,Esser2001}. As we shall demonstrate, that theory corresponds to a weak-interaction limit of the low doping/high temperature version of our polaron theory. To see this, we take the %
limit of a small self-energy in the Dyson equation in Eq.~\eqref{eq:dressedX} as follows:
\begin{align} \label{eq:dysonweak}
    G_X(\omega)\simeq \frac1\omega+\frac1{\omega^2}\Sigma(\omega) \; .
\end{align}
Since the $1/\omega$ terms only have a pole at $\omega=0$ (i.e., at the bare exciton energy), the attractive branch is obtained from the imaginary part of the self-energy. The detailed balance equation~\eqref{eq:PL_func} then implies that the corresponding PL from the attractive branch is
\begin{align}
    P_{A}(\omega)&\simeq -\frac1\pi\frac{Z_0}{Z_{int}} \frac{e^{-\beta \omega}}{\omega^2}\im \Sigma_A(\omega) \nn \\
    &\hspace{-10mm}\simeq \frac{Z_0}{Z_{int}} \frac z\area\sum_\q e^{-\beta(\omega+\epsilon_\q)}|\tilde{\eta}_{\q_r}^{(\0)}|^2\delta(\omega+\epsilon_{r\q_r}+\varepsilon_T)\nn \\
    &\hspace{-10mm}\simeq \frac{Z_0}{Z_{int}} z\, e^{\tfrac{\beta(m\omega +m_T\varepsilon_T)}{m_X}}\left|\tilde{\eta}_{\sqrt{2m_r|\omega+\varepsilon_T|}}^{(\0)}\right|^2\Theta(-\omega-\varepsilon_T)
    \;,
\end{align}
where we used Eq.~\eqref{eq:SigmaTrionWF}. Up to frequency-independent prefactors, this precisely matches the result of Esser \textit{et al.} Thus, the trion-picture PL is already contained within the polaron picture. However, as we have already argued, the trion picture calculation assumes that we are in a weakly interacting limit where the self-energy is much smaller than the frequency. This assumption manifestly breaks down close to the onset of the attractive branch, where the real part of the self-energy diverges, and hence the previous trion picture calculation fails to correctly describe the onset and shape of the attractive branch. On the other hand, it correctly identifies the exponential tail of the PL, which is dominated by the imaginary part of the self-energy. 

For the repulsive branch, the trion picture again uses the weak-interaction limit of the Dyson equation~\eqref{eq:dysonweak}, this time neglecting even the second term on the right hand side. This results in
\begin{align}
    P_{R}(\omega)\simeq \frac{Z_0}{Z_{int}}\delta(\omega),
\end{align}
with a small correction that counteracts the oscillator strength transferred to the attractive branch~\cite{Esser2001}. We see that the trion picture fails to describe the Lorentzian shape of the repulsive polaron, which arises from the exciton-electron scattering states as well as the full Dyson series.

\section{Strong light-matter coupling}
\label{sec:strong-c}
\begin{figure*}
    \centering
    \includegraphics[width=0.85\textwidth]{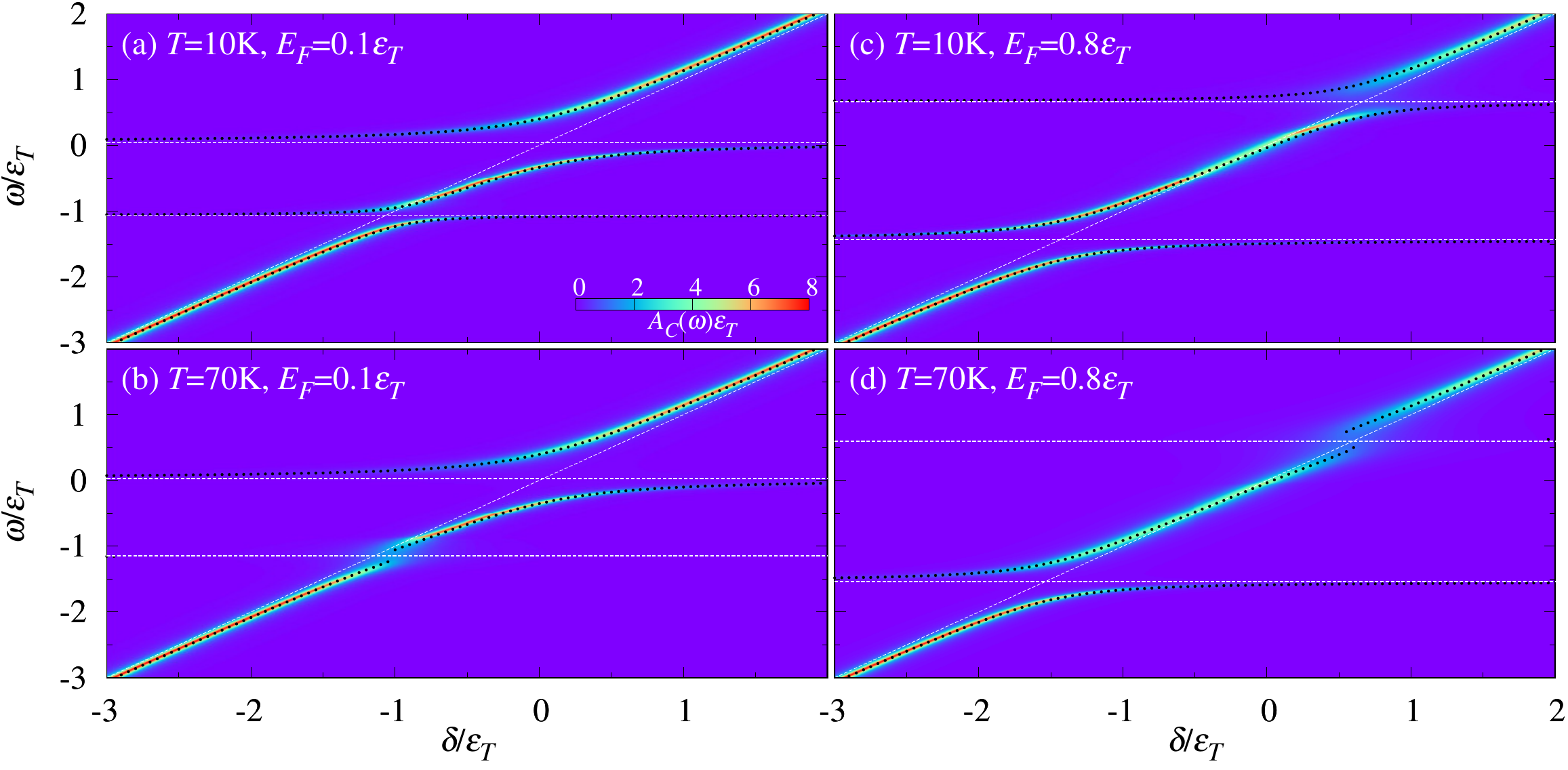}
    \caption{Energy and detuning dependence of the photon spectral function $A_C(\omega)$ in the light-matter strongly coupled system describing a TMD monolayer with an electron doping $E_F=0.1\varepsilon_T$ (panel (a) and (b)) or $E_F=0.8\varepsilon_T$ (panel (c) and (d)), embedded in a microcavity at a finite temperature $T=10K\simeq 0.034 \varepsilon_T$ (panel (a) and (c)) and $T=70K\simeq 0.24\varepsilon_T$ (panel (b) and (d)). The black dots are the LP, MP and UP branches extracted from the three coupled oscillator model (see text). The Rabi splitting is $\Omega=20$meV$\simeq0.8\varepsilon_T$.%
    Here, we have used a broadening of $\eta_C=1$meV$\simeq 0.04\varepsilon_T$ for the photon and $\eta_{X}=0.2\eta_C$ for the matter component.}
    \label{fig:Strong_vs_delta}
\end{figure*}
We now extend the formalism of Sec.~\ref{sec:polaron} to describe recent experiments where a doped TMD monolayer was embedded into a microcavity~\cite{Sidler_NatPhys_2017,Chakraborty_NanoLett2018,Emmanuele_NatComm2020,Koksal_PRR2021,Lyons_NatPh2022}. In this case, the strong coupling between light and matter can lead to the formation of Fermi polaron-polaritons~\cite{Baeten_PRB2015,Sidler_NatPhys_2017}. A particularly important question is whether there still exists strong coupling to light either because of temperature induced broadening effects or because the attractive branch does not correspond to a well-defined quasiparticle.

To describe the light-matter coupled system, we add two terms describing cavity photons and the photon-exciton interaction to the Hamiltonian~\eqref{eq:Hamiltonian}:
\begin{subequations}
\label{eq:Hamiltonian_sc}
\begin{align}
    \hat{H} &= \hat{H}_0 + \hat{H}_{0X}+ \hat{H}_{0C}+ \hat{H}_{int}+\hat{H}_{XC} \\
    \hat{H}_{0C}&=
    \sum_{\k}  \epsilon_{C\k}\hat{a}^{\dag}_{\k}\hat{a}^{}_{\k}
    \\
    \label{eq:HXC}    \hat{H}_{XC}&=\frac{\Omega}{2}\sum_{\k}\left( \hat{x}^{\dag}_{\k}\hat{a}^{}_{\k} + \text{H.c.} \right)\; .
\end{align}
\end{subequations}
Photons are described by the bosonic creation operator $\hat{a}^{\dag}_{\k}$ and, in a cavity, they acquire a quadratic dispersion $\epsilon_{C\k} = \delta+ \k^2/2m_C$~\cite{Microcavities}, where $\delta$ is the photon-exciton energy detuning. Typically the photon mass in a planar microcavity is $m_C = 10^{-5} m_X$~\cite{Microcavities}. The term $\hat{H}_{XC}$~\eqref{eq:HXC} describes an exciton recombining to emit a photon and vice versa, with a coupling strength given by the Rabi splitting $\Omega$~\cite{Microcavities}. In monolayer TMDs, this is typically in the range of 1-2 $\varepsilon_T$~\cite{liu2015strong,dufferwiel2015exciton}.

In order to derive the photon Green's function and the optical absorption in the strong coupling regime, 
one can follow the same procedure employed in Sec.~\ref{sec:polaron}. %
The difference is  that now we formulate a variational ansatz for the time-dependent photon operator, %
with an analogous form to~\eqref{eq:var_imp-op}:
\begin{equation}
    \hat{a}_{\0}^{}(t) \simeq \alpha_{0}^{}(t) \hat{a}_{\0}^{}  + \frac{1}{\area} \sum_{\k,\q} \varphi_{\k\q}^{}(t)  \hat{c}_{\q}^{\dag} \hat{c}_{\k}^{} \hat{x}_{\q-\k}^{}
    + \varphi_{0}^{}(t) \hat{x}_{\0}^{} \; .
\label{eq:var_imp-op_ph}    
\end{equation}
We neglect the dressing of the photon operator by a particle-hole excitation $\frac{1}{\area} \sum_{\k,\q} \alpha_{\k\q}^{}(t)  \hat{c}_{\q}^{\dag} \hat{c}_{\k}^{} \hat{a}_{\q-\k}^{}$. This term involves photon recoil, and therefore implies energies far off resonance from the exciton and trion energies because of the extremely small mass of the photon. 

To obtain the eigenvalue problem for the light-matter coupled system, we introduce the error  operator corresponding to the photon $\hat{e}(t) = i\partial_t \hat{a}_{\0}^{}(t) - [\hat{a}_{\0}^{}(t),\hat{H}]$ and minimize the error function, Eq.~\eqref{eq:error}, with  respect to the variational coefficients $\alpha_{0}^{*}(t)$, $\varphi_{\k\q}^{*}(t)$, and $\varphi_{0}^{*}(t)$. 
Considering the stationary solutions, we find
\begin{subequations}
\label{eq:P3_finiteTph}
\begin{align} \label{eq:P3_1ph}
    E\varphi_{0}^{} &= %
    \Frac{\Omega}{2} \alpha_0^{}  %
-\Frac{v}{\area^2}\sum_{\k,\q} f_\q(1-f_\k) \varphi_{\k\q}^{}
 \\  \label{eq:P3_2ph}  
    E\varphi_{\k\q}^{} &= E_{X\k\q} \varphi_{\k\q}^{} - v\varphi_{0}^{} %
-\Frac{v}{\area}\sum_{\k'}(1-f_{\k'}) \varphi_{\k'\q}^{} %
\\
 \label{eq:P3_3ph}  
    E\alpha_{0}^{} &= \delta \alpha_{0}^{} + \Frac{\Omega}{2} \varphi_{0}^{}\; ,
\end{align}
\end{subequations}
where we have again neglected terms that vanish in the limit $\Lambda\to\infty.$
These equations reduce to those for %
the exciton polaron, Eq.~\eqref{eq:P3_finiteT}, when we take $\Omega=0$ and  $\alpha_0^{}=0$. By following an analogous derivation to that in Sec.~\ref{sec:polaron}, one can easily demonstrate that the retarded photon Green's function in the frequency domain is given by:
\begin{equation}
  G_{C} (\omega) = \sum_n
  \frac{|\alpha_0^{(n)}|^2}{\omega-E^{(n)}+i0} \; .
\end{equation}

For the same reason that we can neglect the particle-hole dressing of the photon operator in the ansatz~\eqref{eq:var_imp-op_ph}, the expressions for the exciton self-energy $\Sigma(\omega)$ in Eqs.~\eqref{eq:Xself-energy} and~\eqref{eq:invT-matrix} are not affected by the  coupling to light. Therefore, we can derive the coupled exciton and photon Green's functions in the strong coupling regime by inverting the matrix~\cite{Levinsen2019}
\begin{equation*}
    \mathbb{G}(\omega) = \begin{pmatrix}
  \omega  -\Sigma(\omega) &  -\Omega/2  \\
  -\Omega/2 &  \omega-\delta \end{pmatrix}^{-1}  \; ,
\end{equation*}
and evaluating the diagonal elements, giving:
\begin{subequations}
\label{eq:dressed_ph}
\begin{align}
\label{eq:strong-c_X}
    G_{X}(\omega) &=\mathbb{G}_{11}(\omega)=\frac{1}{\omega - \Sigma(\omega) - \frac{(\Omega/2)^{2}}{\omega-\delta}}\\
    G_{C}(\omega) &=\mathbb{G}_{22}(\omega)=\frac{1}{\omega-\delta-\frac{(\Omega/2)^{2}}{\omega-\Sigma(\omega)}}
    \; .
\label{eq:dressedX_ph}
\end{align}
\end{subequations}
The expression~\eqref{eq:strong-c_X} for the  exciton Green's function in terms of the self-energy  can also be obtained by following the same procedure as in Appendix~\ref{app:Xself-en}, where we have first eliminated the photon amplitude by solving Eq.~\eqref{eq:P3_3ph}, i.e., $\alpha_0 = \frac{\Omega}{2} \varphi_0^{} (E-\delta)^{-1}$.
Finally, in the strong coupling regime, optical absorption coincides (up to a frequency-independent prefactor) with the photon spectral function~\cite{Cwik_PRA2016}: 
\begin{equation}
    A_{C}(\omega)=-\frac{1}{\pi}\Im G_C(\omega) \; ,
\label{eq:C_Spectr_func}
\end{equation}
which we will focus on in the following.

We plot in Fig.~\ref{fig:Strong_vs_delta} the finite-temperature photon spectral function at normal incidence as a function of energy and detuning $\delta$. When %
$T \ll E_F \lesssim \Omega$, the strong coupling to light leads to three polariton branches, the lower (LP), middle (MP), and upper polariton (UP), as can be seen in Fig.~\ref{fig:Strong_vs_delta}(a,c). %

The existence of %
these polariton modes is connected to the attractive and repulsive exciton branches in the presence of doping. We can capture the behavior of the Fermi polaron-polaritons 
by employing a three coupled oscillator model, which yields the following simplified expression for the photon Green's function
\begin{align}
    \tilde{G}_C(\omega) &= \left.(\omega - \tilde{H})^{-1} \right|_{11}\\
    \tilde{H} & =\begin{pmatrix} \delta-i \eta_C & \Omega_A/2 &
   \Omega_R/2 \\ \Omega_A/2 & E_A - i\Gamma_A & 0
   \\ \Omega_R/2 & 0 & E_R - i\Gamma_R
  \end{pmatrix} \; .
\label{eq:3oscmodel}  
\end{align}
Here, we have explicitly included a cavity photon lifetime $1/\eta_C$ and we have used the extracted parameters for the exciton-polaron branches 
in the absence of coupling to light, namely the energies $E_{A,R}$, spectral weights $Z_{A,R} = (\Omega_{A,R}/\Omega)^2$, and half linewidths $\Gamma_{A,R}$. %
Note that we cannot evaluate the polariton branch energies as the (complex) eigenvalues of the matrix $\tilde{H}$~\cite{Savona_SSC95}. Rather, we have to evaluate first the photon spectral function~\eqref{eq:C_Spectr_func} and then determine the polariton energies from the photon spectral function peak positions. %
The comparison in Fig.~\ref{fig:Strong_vs_delta} between the LP, MP and UP energies obtained in this way and the full calculation demonstrates essentially perfect agreement.

Remarkably, we find that a strong enough light-matter coupling can produce well-defined polariton quasiparticles (where $\Re \tilde{G}_{C}^{-1} (\omega)=0$) for the lower and middle branches  even when there is no attractive polaron quasiparticle and  Eq.~\eqref{eq:GX_poles} is not satisfied. 
However, once the linewidths $2\Gamma_{A,R}$ approach $\Omega_{A,R}$, the energy splitting between branches closes, indicating a loss of strong coupling. We observe this in the low-doping regime for LP and MP
[Fig.~\ref{fig:Strong_vs_delta}(b)],
 and in the high-doping regime for MP and UP [Fig.~\ref{fig:Strong_vs_delta}(c,d)].
Within the three coupled oscillator model, this transition from weak to strong light-matter coupling approximately occurs when $2\Gamma_{A,R} \gtrsim \Omega_{A,R}$, as expected, though there are small deviations when the attractive branch is no longer a well-defined polaron quasiparticle.  
To conclude, the effect of temperature on this transition can be easily accounted for by considering its effect on the quasiparticle linewidths and spectral weights (see Sec.~\ref{sec:results}).

\section{Conclusions and perspectives}
\label{sec:concl}
We have studied the optical properties of a doped two-dimensional semiconductor at a finite temperature using a Fermi-polaron approach involving a single excitation of the fermionic medium. Our results reveal that the attractive branch can experience a smooth transition from a regime where it is a well-defined quasiparticle to a regime where is subsumed into a  broad continuum of trion-hole scattering states. This crossover results in a strong change in the spectral lineshape and can be driven by either decreasing doping or increasing temperature, but it cannot occur at zero temperature. While the Fermi polaron theory is able to capture both limits, theories based on the trion wavefunction necessarily only apply in the limit where there is no a well-defined quasiparticle. In particular, %
we formally show that the trion theory corresponds to %
a weak-interaction limit of our finite-temperature Fermi polaron theory.

In the regime of strong light-matter coupling, we have shown how the temperature can modify the properties of the Fermi polaron-polaritons. 
We demonstrate that the strong-to-weak coupling crossover observed at finite temperature for the attractive branch at low enough doping and the repulsive branch in the high doping regime can be explained in terms of the linewidths and spectral weights of the two branches.

In future studies, it would be interesting to investigate how the quasiparticle transition of the attractive branch, driven by either temperature or doping, would affect the interactions between exciton impurities. In particular, common descriptions of polaron-polaron interactions use Landau's theory of dilute solutions~\cite{Yu2012}, which assumes well-defined quasiparticles. Such interactions could, for instance, be measured using coherent multidimensional spectroscopy on gated 2D materials, similarly to recent experiments on intrinsically doped samples~\cite{Muir2022}. 

\acknowledgments
We gratefully acknowledge useful discussions with Dmitry Efimkin. AT and FMM acknowledge financial support from the Ministerio de Ciencia e Innovaci\'on (MICINN), project No.~AEI/10.13039/501100011033 (2DEnLight).
FMM acknowledges financial support from the Proyecto Sinérgico CAM 2020 Y2020/TCS-6545 (NanoQuCo-CM).
JL and MMP are supported through Australian Research Council Future Fellowships FT160100244 and FT200100619, respectively. JL, BM and MMP also acknowledge support from the Australian Research Council Centre of Excellence in Future Low-Energy Electronics Technologies (CE170100039).

\appendix

\section{Finite impurity momentum}
\label{app:finiteQ}
For completeness, we generalize the finite-temperature polaron formalism illustrated in Sec.~\ref{sec:polaron} to absorption and photoluminescence at finite  %
momentum. %
This %
can %
in principle be measured in doped semiconductors using angle-resolved photoemission spectroscopy~\cite{Madeo-Heinz_science2020,Man-Heinz_sciadv2021}.
Similarly  to Eq.~\eqref{eq:var_imp-op}, we approximate the exciton operator in the Heisenberg picture at finite %
momentum $\hat{x}_{\Q}^{}(t) = e^{i\hat{H}t} \hat{x}_{\Q}^{} e^{-i\hat{H}t}$ as
\begin{multline}
    \hat{x}_{\Q}^{}(t) \simeq \varphi_{0}^{(\Q)}(t) \hat{x}_{\Q}^{} + \frac{1}{\area} \sum_{\k,\q} \varphi_{\k\q}^{(\Q)}(t)  \hat{c}_{\q}^{\dag} \hat{c}_{\k}^{} \hat{x}_{\Q+\q-\k}^{}\; .
\label{eq:var_imp-opQ}    
\end{multline}
The derivation then follows similarly to the zero momentum case. We minimize the error function $\Delta_{\Q}(t) = \langle \hat{e}_{\Q}^{}(t) \hat{e}_{\Q}^{\dag}(t) \rangle_{\beta}$, where  $\hat{e}_{\Q}^{}(t) = i\partial_t \hat{x}_{\Q}^{}(t) - [\hat{x}_{\Q}^{}(t),\hat{H}]$, obtaining the following
eigenvalue problem:
\begin{subequations}
\label{eq:P3_finiteTQ}
\begin{align}
    E^{(\Q)}\varphi_{0}^{(\Q)} &= \epsilon_{X\Q} \varphi_{0}^{(\Q)}   
    -\Frac{v}{\area^2}\sum_{\k,\q} f_\q(1-f_\k) \varphi_{\k\q}^{(\Q)}
\label{eq:P3_1Q} \\
    E^{(\Q)}\varphi_{\k\q}^{(\Q)} &= E_{X\Q\k\q} \varphi_{\k\q}^{(\Q)} - v\varphi_{0}^{(\Q)} %
    -\Frac{v}{\area}\sum_{\k'}(1-f_{\k'}) \varphi_{\k'\q}^{(\Q)} %
    \; ,
 \label{eq:P3_2Q}   
\end{align}
\end{subequations}
where $E_{X\Q\k\q} = \epsilon_{X\Q+\q-\k} + \epsilon_\k - \epsilon_\q$. The exciton Green's function can thus be written in terms of the eigenvalues $E^{(\Q,n)}$ and eigenvectors $\varphi_{0}^{(\Q,n)}$ as
\begin{equation}
  G_{X} (\Q, \omega) = \sum_n
  \frac{|\varphi_0^{(\Q,n)}|^2}{\omega-E^{(\Q,n)}+i0} \; .
\end{equation}

Equivalently, the  exciton Green's function can be written in terms of the exciton self-energy at finite momentum
\begin{align}
    G_{X}(\Q,\omega) &=\frac{1}{\omega - \epsilon_{X\Q} - \Sigma(\Q,\omega)}\\
    \Sigma(\Q,\omega) &=\frac{1}{\area}\sum_{\q} f_\q \mathcal{T}(\q+\Q, \omega+\epsilon_{\q})\; ,
\label{eq:Xself-energy_Q}
\end{align}
where the inverse of the $T$ matrix is defined in Eq.~\eqref{eq:invT-matrix}. This follows the same procedure as in the case of $\Q=\0$, which is described in Appendix~\ref{app:Xself-en}.

The absorption of a photon with momentum $\Q$ is given by
\begin{equation}
    A(\Q,\omega) =-\frac{1}{\pi}\Im G_X(\Q,\omega) \; .
\end{equation}
Absorption and photoluminescence can be connected using detailed balance conditions as derived in the main text, starting from the Fermi's golden rule definitions:
\begin{subequations}
\begin{align}
    A(\Q,\omega) &= \sum_{n,\nu} \langle n| \hat{\rho}_{0} |n \rangle |\langle \nu| \hat{x}_{\Q}^\dag |n \rangle |^2 \delta (E_{\nu n %
    } - \omega)
\label{eq:FermisGR_A_Q}\\
    P(\Q,\omega) &= \sum_{n,\nu} \langle \nu| \hat{\rho} |\nu \rangle |\langle n| \hat{x}_{\Q}^{} |\nu \rangle |^2 \delta (E_{\nu n %
    } - \omega) \; .
\label{eq:FermisGR_PL_Q}    
\end{align}
\end{subequations}
Thus, the detailed balance condition is identical to that at zero momentum:
\begin{equation}
    P(\Q,\omega) = \Frac{Z_0}{Z_{int}} e^{-\beta\omega%
    } A(\Q,\omega) \; .
\label{eq:PL_func_Q}
\end{equation}
\section{Exciton self-energy}
\label{app:Xself-en}
In this appendix, we demonstrate that the exciton self-energy can be derived starting from the eigenvalue equations~\eqref{eq:P3_finiteT} and that it coincides with the expression~\eqref{eq:Xself-energy} that can be derived within a $T$ matrix formalism.
Let us start from the eigenvalue problem~\eqref{eq:P3_finiteT}
which can be
rewritten in terms of the auxiliary function $\chi_{\q}^{}$:
\begin{subequations}
\label{eq:P3_chi}
\begin{align}
\label{eq:P3_chi_1}
    \chi_{\q}^{} &=%
    \frac{v}{\area}\sum_{\k} (1-f_\k) \varphi_{\k\q}^{}\\
    \label{eq:P3_chi_2}
    E \varphi_{0}^{} &=-\frac{1}{\area}\sum_{\q} f_\q \chi_{\q}^{}\\
    E \varphi_{\k\q}^{} &=
    E_{\text{X}\k\q} \varphi_{\k\q}^{}-v\varphi_0- \chi_{\q}^{} \; .
\label{eq:P3_chi_3}    
\end{align}
\end{subequations}
Introducing Eq.~\eqref{eq:P3_chi_3} into Eq.~\eqref{eq:P3_chi_1} and solving for $\chi_\q$ we obtain
\begin{equation}
    \chi_{\q}^{} = \varphi_{0}^{} \left[\frac{1}{v}+ 
    \frac{1}{\area}\sum_{\k}  \frac{1-f_\k}{E-E_{X\k\q}} \right]^{-1} \; ,
\label{eq:chi_sol}    
\end{equation}
where we have used that $(v/\area) \sum_\k (1-f_{\k})/(E-E_{X\k\q}) \to -1$ in the limit $\Lambda\to\infty$.
Substituting~\eqref{eq:chi_sol} into Eq.~\eqref{eq:P3_chi_2}, one thus finally obtains an implicit equation for the energy $E$:
\begin{equation}
    E =  - \frac{1}{\area}\sum_{\q} f_\q \left[\frac{1}{v} +\frac{1}{\area}\sum_{\k}
  \frac{1-f_\k}{E-E_{X\k\q}} \right]^{-1} \; .
\end{equation}
This expression coincides with the pole of the exciton Green's function~\eqref{eq:dressedX} and therefore we can identify the self-energy term correcting the exciton energy because of the exciton-electron interaction as the term:
\begin{equation}
    \Sigma(E) \equiv - \frac{1}{\area}\sum_{\q} f_\q \left[\frac{1}{v} +\frac{1}{\area}\sum_{\k}
  \frac{1-f_\k}{E -E_{X\k\q}} \right]^{-1} \; .
\label{eq:def_self-en}    
\end{equation}
Recalling that $E_{X\k\q} = \epsilon_{X\q-\k} + \epsilon_\k-\epsilon_\q$, we obtain exactly the expression~\eqref{eq:Xself-energy} in the main text. By a completely equivalent procedure, we can arrive at the self-energy in the case of finite exciton momentum, Eq.~\eqref{eq:Xself-energy_Q}.

\section{Trion}
\label{app:trion}
In this appendix, we summarize for completeness the known trion properties within our model Hamiltonian~\eqref{eq:Hamiltonian}.
At zero temperature, a trion with %
momentum 
$\Q$ in the presence of a Fermi sea $|FS\rangle = \prod_{|\q|<k_F} c_{\q}^\dag |0\rangle$ is described as:
\begin{equation}
    |T_2^{(\Q)}\rangle= \Frac{1}{\sqrt{\area}}\sum_{|\k|>k_F} \eta_{\k}^{(\Q)} \hat{x}^{\dag}_{\Q-\k} \hat{c}_{\k}^\dag | FS\rangle \; .
\label{eq:FS_trion}
\end{equation}
The trion wave function $\eta_{\k}^{(\Q)}$ satisfies the Schr\"odinger equation~\cite{Parish_PRA11}
\begin{equation}
  E_T^{(\Q,E_F)} \eta_{\k}^{(\Q)} = \left(\epsilon_{X\Q-\k} + \epsilon_{\k}\right) \eta_{\k}^{(\Q)} - \frac{v}{\area}\sum_{|\k'|>k_F}^{\Lambda} \eta_{\k'}^{(\Q)} \; ,
\label{eq:trion_eq}  
\end{equation}
and the trion energy can be evaluated by solving the implicit equation:
\begin{equation}
    \Frac{1}{v} =  \Frac{1}{\area}\sum_{|\k|>k_F}^{\Lambda} \Frac{1}{-E_T^{(\Q,E_F)} + \epsilon_{X\Q-\k} + \epsilon_\k} \; . 
\label{eq:trion_finiteQ-EF}    
\end{equation}
Note that some care has to be taken when comparing the trion energy with that of a bare exciton, since the Fermi sea in Eq.~\eqref{eq:FS_trion} involves one less electron. This difference has been taken into account in Eqs.~\eqref{eq:Eplus} and \eqref{eq:Eminus} of the main text.
We now discuss the solution of Eq.~\eqref{eq:trion_finiteQ-EF} in various limits.

When $E_F=0$, it is profitable to introduce the relative momentum in the center of mass frame $\q_r$, so that the electron momentum becomes $\k = \q_r + \Q_c$, with $\Q_c=\frac{m}{m_T} \Q$, where $m_T = m_X + m$ is the trion mass, and the exciton momentum becomes $\Q-\k = \Q_X - \q_r$, with  $\Q_X= \frac{m_X}{m_T} \Q$. Now one has that relative and center of mass kinetic energies factorize, $\epsilon_{X\Q-\k} + \epsilon_{\k} = \epsilon_{X\Q_X-\q_r} + \epsilon_{\q_r+\Q_c} = \epsilon_{r\q_r} + \epsilon_{T\Q}$, where $\epsilon_{r\q_r}  = \frac{q_r^2}{2m_r}$, $m_r = m m_X/m_T$ is the trion reduced mass, and $\epsilon_{T\Q} = \frac{Q^2}{2m_T}$. Thus, the trion energy and wave function are given by
\begin{align}
    E_T^{(\Q,E_F=0)} &= -\varepsilon_T + \epsilon_{T\Q} \\
    \tilde{\eta}_{\q_r}^{(\Q)} &= \Frac{\sqrt{Z_T}}{\varepsilon_T + \epsilon_{r\q_r}} = \tilde{\eta}_{\q_r}^{(\0)} \; ,
\label{eq:trion-vacuum_Q}
\end{align}
where $Z_T = 2\pi \varepsilon_T/m_r$ and where $\tilde{\eta}_{\q_r}^{(\Q)} = \eta_{\q_r + \Q_c}^{(\Q)}$.

At finite doping and zero %
momentum $\Q=\0$, Eqs.~\eqref{eq:trion_eq} and~\eqref{eq:trion_finiteQ-EF} can also be solved exactly to give %
\begin{align}
\label{eq:trion-EF_Q01}
    E_T^{(\0,E_F)} &= - \varepsilon_T+\frac{m_T}{m_X}E_F\\
    \eta_{\k}^{(\0)} &=  \frac{\sqrt{Z_T}}{\varepsilon_T - \frac{m_T}{m_X} E_F + \epsilon_{r\k}} \; .
\label{eq:trion-EF_Q0}
\end{align}

At finite doping and %
finite momentum,  Eq.~\eqref{eq:trion_finiteQ-EF} can be solved analytically~\cite{Parish_PRA2013} %
to find an implicit equation for $E_T^{(\Q,E_F)}$. %
Here, one finds that the trion acquires a finite %
momentum %
when $E_F > \frac{m_X m_r}{m^2} \varepsilon_T$. %

We can now analyze the trion coupling strength to light. At zero doping, $E_F=0$, trions do not couple to light, and thus cannot be  probed optically. This is because the matrix  element between a trion state and a cavity photon plus a majority particle at zero momentum $|C+1\rangle = \hat{a}_\0^\dag \hat{c}_\0^\dag |0 \rangle$ 
of the light-matter interaction term $\sum_{\k} \hat{x}^{\dag}_{\k}\hat{a}^{}_{\k}$ is given by
\begin{equation}
    \langle T_2^{(\0)} | \sum_{\k} \hat{x}^{\dag}_{\k}\hat{a}^{}_{\k} |C+1\rangle = \Frac{1}{\sqrt{\area}} \eta_{\0}^{(\0)}\; .
\end{equation}
Taking the squared amplitude of this matrix element, we see that the coupling to light of a single trion scales as $1/\area$, which is vanishingly small.

On the other hand, if we have $N$ electrons within $\area$, then the coupling to light scales instead as $N/\area=n\sim E_F$. Thus, even though the coupling per electron vanishes, the net effect of having an electronic medium is to create a continuum of states, with a total spectral weight that scales as $E_F$, in agreement with Ref.~\cite{GlazovJCP2020}.

\section{Numerical evaluation of the many-body $T$ matrix and the exciton self-energy}
\label{app:mb_Tmatrix}
In contrast to the vacuum $T$ matrix~\eqref{eq:2b_Tmatrix},  the many-body correction in Eq.~\eqref{eq:mb_Tmatrix} cannot be evaluated analytically. The approach we illustrate in the following allows us to treat it in a simpler way and to evaluate it numerically, even when the intrinsic exciton linewidth  $2\eta$ is set to zero. Let us start by re-writing~\eqref{eq:mb_Tmatrix} in an equivalent form:
\begin{multline}
    \Pi_{mb}(\q,\omega) = - \frac{1}{\area} \sum_{\k} \Frac{f_{\k + \frac{m}{m_T}\q}}{\omega - \epsilon_{\k+\frac{m}{m_T}\q} - \epsilon_{X\k-\frac{m_X}{m_T}\q} + i0^+} \\= - \int \frac{dkk}{2\pi} \frac{\int\frac{d\theta}{2\pi} f_{\k + \frac{m}{m_T}\q}}{\omega - \epsilon_\q \frac{m}{m_T} - \epsilon_\k \frac{m_T}{m_X} + i0^+}\;.
\end{multline}
The $k$-integral can then be evaluated numerically by applying the Sokhotski–Plemelj theorem
\begin{align*}
\int dx\frac{F(x)}{x + i0^+}&=- i\pi F(0)+\mathcal{P}\int dx \frac{F(x)}{x} \;,
\end{align*}
where $\mathcal{P}[\dots]$ is the integral principal part.

As far as the $\q$-integral for evaluating the exciton self-energy~\eqref{eq:Xself-energy} is concerned, it can be re-written in the following equivalent form by defining $y=q^2$:
\begin{equation*}
    \Sigma(\omega) =\int\frac{dy}{4\pi} \frac{f_{\sqrt{y}}}{ \mathcal{T}^{-1} (\sqrt{y} , \omega + \epsilon_{\sqrt{y}})} \; .
\end{equation*}
This integral has a pole when $\Re \mathcal{T}^{-1} = 0 = \Im \mathcal{T}^{-1}$. Using a model involving
contact interactions in 2D means that such a pole always
exists, and furthermore it is a simple pole~\cite{Engelbrecht1992}. Thus, we define $y^*= y^*(\omega)$ to be the pole of the $T$ matrix.
 
In this case, we can apply the Cauchy residue theorem to write:
\begin{widetext}
\begin{multline}
    \Sigma(\omega) = - i\pi\; \text{sign}\left[\Im \mathcal{T}^{-1} (\sqrt{y^*}, \omega + \epsilon_{\sqrt{y^*}})\right] \frac{\Theta(y^*)}{4\pi}  \mathcal{R}es \left[\frac{f_{\sqrt{y}}}{\Re\left[\mathcal{T}^{-1}(\sqrt{y},\omega + \epsilon_{\sqrt{y}})\right]}\right]_{y^*}\\
    +  \mathcal{P} \int\frac{dy}{4\pi} \frac{f_{\sqrt{y}}}{\Re \mathcal{T}^{-1}(\sqrt{y},\omega+\epsilon_{\sqrt{y}}) + i\Im\mathcal{T}^{-1} (\sqrt{y},\omega + \epsilon_{\sqrt{y}})} \; ,
\label{eq:Princ_Sigma}
\end{multline}
\end{widetext}
where the principal part prescription is used in the vicinity of the pole at $y^*$, and the residue can be evaluated as
\begin{multline}
    \mathcal{R}es \left[\frac{f_{\sqrt{y}}}{\Re\left[\mathcal{T}^{-1}(\sqrt{y},\omega + \epsilon_{\sqrt{y}})\right]}\right]_{y^*}\\
    = \Frac{f_{\sqrt{y^*}}}{\displaystyle\left| \frac{\partial \Re \mathcal{T}^{-1}(\sqrt{y},\omega+\epsilon_{\sqrt{y}}) }{\partial y}\right|_{y^*}}\; .
\end{multline}

From the expression of the exciton self-energy, we can get the exciton Green's function~\eqref{eq:dressedX}, absorption~\eqref{eq:X_Spectr_func}, and photoluminesence~\eqref{eq:PL_func}. These results allow us to evaluate the absorption and photoluminescence without any intrinsic homogeneous broadening for the exciton. For absorption alone, the integration method illustrated here is unnecessary, as one can conveniently shift the frequency to the complex plane, $\omega \mapsto \omega + i\eta$, where $\eta$ is related to the exciton decay rate, giving well converged results. However, as explained in the main text, for the luminescence one has to adopt the numerical method illustrated here. Thus, in order to compare with experiments, we introduce the effect of  lifetime broadening at the end by convolving the photoluminescence with a Lorentzian profile with broadening $\eta$:
\begin{equation}
    \bar{P}(\omega') = \int_{-\infty}^{\infty} d\omega' P(\omega') \Frac{1}{\pi} \Frac{\eta}{(\omega - \omega')^2 + \eta^2}\; .
\label{eq:Lorentz_conv}
\end{equation}

\begin{figure}
    \centering
    \includegraphics[width=1.0\columnwidth]{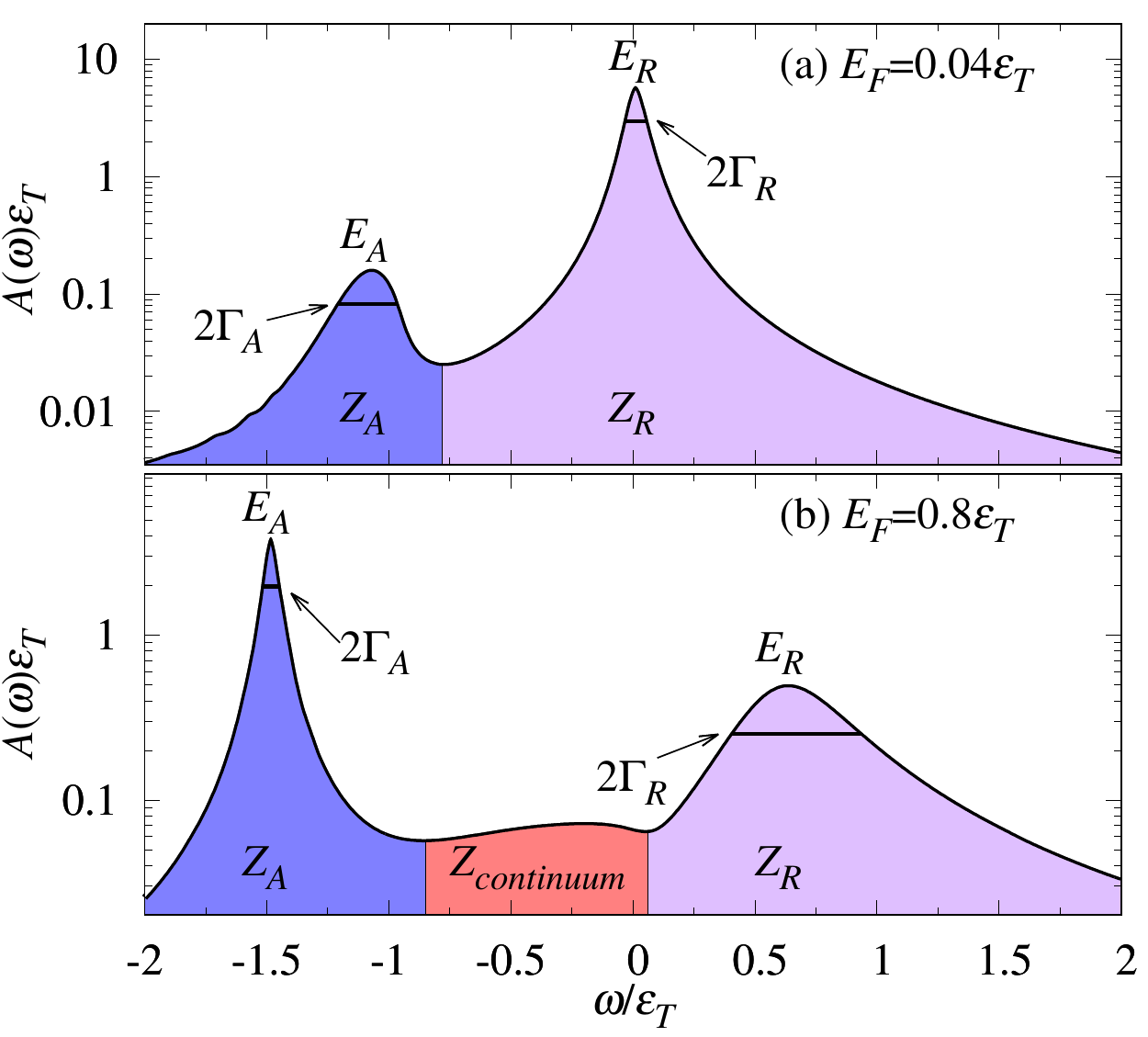}
    \caption{Spectral function $A(\omega)$ at  temperature $T=50$~K$\simeq 0.17\varepsilon_T$ for two different dopings: $E_F= 0.04\varepsilon_T$ (a) and $E_F=0.8 \varepsilon_T$ (b). From the spectral function profile we extract the attractive and repulsive polaron energy $E_{A,R}$ as the peak position, the 
    linewidth $2\Gamma_{A,R}$ as the peak FWHM, and the polaron spectral weight $Z_{A,R,continuum}$ as the area under the peak. 
    While at low doping in panel (a) the trion-hole continuum is merged with the attractive branch (which, here, ceases to be a quasiparticle resonance) and $Z_{continuum} = 0$, at larger doping in panel (b) the polaron resonances are separated by the trion-hole continuum with a finite spectral weight $Z_{continuum} \ne 0$.}
\label{fig:eval_w-Z-Gamma}
\end{figure}

\section{Extracting polaron properties at finite temperature}
\label{app:properties}
In this appendix we illustrate how to extract  the polaron energies, spectral weights, and half linewidths of Figs.~\ref{fig:T50_vs_T0} and ~\ref{fig:EF0.4_Properties} from the spectral function profile $A(\omega)$. This is illustrated in Fig.~\ref{fig:eval_w-Z-Gamma}. The polaron energy $E_{A,R}$ is evaluated as the spectral function peak position, the linewidth $2\Gamma_{A,R}$ as the full width at half maximum (FWHM) and the spectral weight as the area under the peak. We have used the location of the spectral function minima as limits for the integrals evaluating the spectral weights: If there is only one minimum, this is the upper (lower) bound for evaluating $Z_A$ ($Z_R$), while if there are two minima, these are the limits for evaluating the trion-hole continuum spectral weight $Z_{continuum}$. This criterion is the origin of the discontinuity of the residues in Figs.~\ref{fig:T50_vs_T0}, \ref{fig:EF0.4_Properties}, and \ref{fig:compare_FL-th}.

Note that, in the cases shown in Fig.~\ref{fig:eval_w-Z-Gamma}, at low doping of panel (a) the trion-hole continuum does not appear because is merged with the attractive branch, which in this case, ceases to be a polaron quasiparticle resonance in the sense explained in Sec.~\ref{sec:no-QP}. This can also be appreciated by the non-Lorentzian and asymmetric form of the spectral function around $E_A$. By contrast, at larger doping in panel (b), the attractive branch is separated from the continuum, is a polaron quasiparticle in this case and recovers the Lorentzian symmetric shape.

\section{Relationship between electron-exciton $T$ matrix and the trion wave function}
\label{app:Tfromwave}
While Eq.~\eqref{eq:SigmaTrionWF} was derived for the special case of contact exciton-electron interactions, it is in fact general and can be applied for any realistic trion wave function. To see this, we note that the generalization of Eq.~\eqref{eq:SEsmallz} to arbitrary two-body transition operator $\hat T_0$ is
\begin{align}
    \Sigma(\omega)\simeq \frac z\area\sum_\q e^{-\beta\epsilon_\q} \expval{\hat T_0(\omega+\epsilon_\q)}{\Q,\q_r}\;.
    \label{eq:SEsmallzGeneral}
\end{align}
We have defined the two-body state $\ket{\Q,\q_r}\equiv \hat c_\q^\dag x_\0^\dag\ket{vac}$ in terms of the relative momentum $\q_r$ and the total %
momentum $\Q$ which are related to the electron momentum via $\q_r+\Q m/m_T=\q$ and the exciton momentum via $-\q_r+\Q m_X/m_T=\0$.

To evaluate the matrix element of the transition operator for an electron-exciton interaction that yields a trion bound state, we use the relationship between the Green's operator and the transition operator at energy $E$:
\begin{align}\label{eq:2partgreen}
    \hat G(E) = \hat G_0(E)+\hat G_0(E)\hat T(E)\hat G_0(E)\;.
\end{align}
Here, $\hat G(E)=\frac1{E-\hat H+i0}$ and $\hat G_0(E)=\frac1{E-\hat H_0-\hat H_{0X}+i0}$ (which, for the two-body problem, should be evaluated in the canonical ensemble, effectively taking $\mu=0$). Close to the trion resonance, we can neglect the first term, and we therefore have the spectral representation
\begin{align}
    \expval{\hat T_0(E)}{\Q,\q_r}\hspace{-10mm}&\nn\\
    &\simeq \expval{\hat G_0^{-1}(E)\hat G(E)\hat G_0^{-1}(E)}{\Q,\q_r} \nn \\
    &=(E-\epsilon_{\q})^2\expval{\hat G(E)}{\Q,\q_r}\nn\\ \label{eq:spectral}
    & \simeq (E-\epsilon_{\q})^2\frac{|\tilde{\eta}_{\q_r}^{(\0)}|^2}{E+\varepsilon_T-\epsilon_{T\Q}+i0}\nn \\
    & \simeq \frac{(-\varepsilon_T+\epsilon_{T\Q}-\epsilon_{\q})^2|\tilde{\eta}_{\q_r}^{(\0)}|^2}{E+\varepsilon_T-\epsilon_{T\Q}+i0}\;,
\end{align}
where in the third step we approximated the expectation value by inserting the trion state, and in the last step we used the pole condition. Taking $E=\omega+\epsilon_\q$ and using $\Q=\q$ we see that Eq.~\eqref{eq:SEsmallzGeneral} reduces to Eq.~\eqref{eq:SigmaTrionWF} which demonstrates that it holds for arbitrary electron-exciton interactions that lead to trion formation.

In the special case of contact interactions, the numerator of Eq.~\eqref{eq:spectral} reduces to $Z_T$ due to Eq.~\eqref{eq:trion-vacuum_Q}, and therefore it reproduces the pole expansion in Eq.~\eqref{eq:Tvacpole} as it should. Note that the fact that the $T$ matrix in this case is independent of the relative momentum is a special feature of contact interactions.

\bibliography{Refs_finiteT.bib}

\end{document}